%% file: qsomode_rev.tex
\documentclass[useAMS,usenatbib,usegraphicx]{mn2e}

\include{journal}
\usepackage{graphicx}
\usepackage[latin1]{inputenc}
\usepackage{color}
\usepackage{times}
\usepackage{natbib}
\usepackage{setspace}
\usepackage{hyperref}
\newif\ifAMStwofonts
\AMStwofontstrue
\definecolor{red}{rgb}{1,0.,0.}

\newcommand{\morgana}{{\sc morgana }}

\newcommand{\msun}{{\rm M}_\odot}
\newcommand{\msunyr}{{\rm M}_\odot\ {\rm yr}^{-1}}
\newcommand{\gaea}{\sc{gaea}}
\newcommand{\ragaeaF}{\sc{f06-gaea}}
\newcommand{\ragaeaH}{\sc{hq11-gaea}}
\def\lesssim{\lower.5ex\hbox{$\; \buildrel < \over \sim \;$}}
\def\gtrsim{\lower.5ex\hbox{$\; \buildrel > \over \sim \;$}}
\title[{\sc AGN in {\gaea}}] {The Rise of Active Galactic Nuclei in the
  GAlaxy Evolution and Assembly semi-analytic model.}
\author[Fontanot et al.]{
  \parbox[t]{\textwidth}{Fabio Fontanot$^{1,2}$\thanks{E-mail:
      fabio.fontanot@inaf.it}, Gabriella De Lucia$^1$, Michaela
    Hirschmann$^3$, Lizhi Xie$^4$, Pierluigi Monaco$^{5,1,2}$, Nicola
    Menci$^6$, Fabrizio Fiore$^1$, Chiara Feruglio$^{1,2}$, Stefano
    Cristiani$^{1,2}$, Francesco Shankar$^7$}
    \vspace*{8pt}\\
    $^1$ INAF - Astronomical Observatory of Trieste, via G.B. Tiepolo 11, I-34143 Trieste, Italy \\
    $^2$ IFPU - Institute for Fundamental Physics of the Universe, via Beirut 2, 34151, Trieste, Italy \\
    $^3$ DARK, Niels Bohr Institute, University of Copenhagen, Lyngbyvej 2, DK-2100 Copenhagen, Denmark \\
    $^4$ Tianjin Astrophysics Center, Tianjin Normal University, Binshuixidao 393, 300384, Tianjin, China\\
    $^5$ Astronomy Sector, Department of Physics, University of Trieste, via Tiepolo 11, 34143, Trieste, Italy\\
    $^6$ INAF - Astronomical Observatory of Rome, via Frascati 33, I-00078, Monteporzio, Italy \\
    $^7$ Department of Physics and Astronomy, University of Southampton, Highfield, SO17 1BJ, UK \\
}

\begin{document}
\date{Accepted ... Received ...}

\maketitle

\begin{abstract} 
  We present a new implementation of the GAlaxy Evolution and Assembly
  ({\gaea}) semi-analytic model, that features an improved modelling
  of the process of cold gas accretion onto supermassive black hole
  (SMBHs), derived from both analytic arguments and high-resolution
  simulations. We consider different scenarios for the loss of angular
  momentum required for the available cold gas to be accreted onto the
  central SMBHs, and we compare different combinations of triggering
  mechanisms, including galaxy mergers and disc instabilities in star
  forming discs. We compare our predictions with the luminosity
  function (LF) observed for Active Galactic Nuclei (AGN) and we
  confirm that a non-instantaneous accretion timescale (either in the
  form of a low-angular momentum reservoir or as an assumed light
  curve evolution) is needed in order to reproduce the measured
  evolution of the AGN-LF and the so-called AGN-downsizing
  trend. Moreover, we also study the impact of AGN feedback, in the
  form of AGN-driven outflows, on the SF properties of model galaxies,
  using prescriptions derived both from empirical studies or from
  numerical experiments. We show that AGN-driven outflows are
  effective in suppressing the residual star formation rate in massive
  galaxies ($> 10^{11} \msun$) without changing their overall assembly
  history. These winds also affect the SFR of lower mass galaxies,
  resulting in a too large fraction of passive galaxies at $< 10^{10}
  \msun$. Finally, we study the Eddington ratio distribution as a
  function of SMBH mass, showing that only objects more massive than
  $10^8 \msun$ are already in a self-regulated state as inferred from
  observations.
\end{abstract}


\section{Introduction}\label{sec:intro}                                       
The role of Active Galactic Nuclei (AGN\footnote{In this paper we will
  define as AGN all sources powered by gas accretion onto SMBHs,
  irrespective of their luminosities. It is custom in the literature
  to refer to the brightest AGN (formally $M_B>-23.5$) as
  Quasi-Stellar Objects or QSOs.}) in the evolution of different
galaxy populations has been at the centre of considerable debate in
the last decade. An increasing consensus on a pivotal role played by
AGN in galaxy evolution has been initially motivated by the existence
of well defined correlations between the mass $M_{\rm BH}$ of the
central super-massive black hole (SMBH) powering the AGN phenomenon
and the properties (either velocity dispersion, luminosity or stellar
mass $M_{\rm bul}$) of the spheroidal component (i.e. bulge) of the
host galaxy \citep{Magorrian98,HaringRix04}. These relations are
commonly referred to as the BH-Bulge relations. At the same time, from
a theoretical point of view, AGN have been proposed as an ideal
solution for a number of long standing problems in galaxy
evolution. Indeed, luminous AGN can release large amounts of thermal
and kinetic energy in a short time, possibly powering large scale
galactic winds \citep{SilkRees98, Fabian99} able to deplete the host
galaxy from its cold gas content and abruptly stop the galaxy star
formation activity \citep[see e.g.][]{DiMatteo05,
  MonacoFontanot05}. Moreover, radio galaxies have been proposed
\citep[see e.g.][]{Croton06, Bower06} as a viable solution to quench
the expected large cooling flows in massive dark matter haloes
(DMHs). In this case, inefficient accretion on the central SMBHs is
believed to drive radio jets, that can efficiently transport the
energy released in the central regions of the host galaxy to the
outskirts of its parent DMH \citep{KaiserBinney03}. One caveat with
this scenario is that the predicted frequency of radio galaxies among
massive galaxies and/or DMHs may be in tension with the observed
fractions \citep[see e.g.][]{Fontanot11a}.

The redshift evolution of the AGN population has been used as a
constraint for theoretical models of galaxy formation \citep[see
  e.g][among others]{Fontanot06, Menci08, Hirschmann12,
  Hirschmann14}. In particular, several authors have used the
so-called ``downsizing'' observed in AGN luminosity functions (LFs),
i.e. the evidence that the space density of luminous AGN peaks at an
earlier redshift than that of fainter sources \citep{Hasinger05,
  LaFranca05}, to understand the relative importance of physical
mechanisms regulating the growth of SMBH via multiphase gas
accretion. The BH-Bulge relation and its redshift evolution have also
been long used as important constraints for the co-evolution of SMBHs
and their host galaxies \citep{KormendyHo13, Decarli10} and actually
also to tune the AGN feedback efficiency. In particular, from a
theoretical point of view, it has been shown that different galaxy
populations may follow a quite different evolution towards the
$z\sim0$ relations, i.e. some of them grow their bulges faster than
their central SMBHs and viceversa, depending on their individual
histories \citep{Lamastra10, Volonteri12}. Moreover, recent studies
have questioned the robustness of the local determination of these
scaling relations both in terms of its overall shape
\citep{GrahamScott15, Fontanot15d} and normalization
\citep{Shankar16}, when additional galaxy populations and selection
biases are taken into account. Moreover, recent statistical analyses
favour velocity dispersion, and not stellar or bulge mass, as the
leading galaxy property connected to BH mass \citep{Shankar19a,
  deNicola19, Marsden20}.

Another critical issue relates to the role of AGN feedback, and in
particular of AGN-driven winds in regulating the star formation rates
(SFRs) of host galaxies. Several observations probe the multiphase
nature of the gaseous outflows and their connection with the central
AGN activity \citep{Feruglio15, Fiore17}. The role of AGN-driven winds
is complex. On the one hand they are supposed to suppress star
formation by removing copious amounts of cold gas from the host galaxy
\citep{Feruglio10}, but there is also some evidence for ``positive''
feedback \citep{Cresci15}, i.e. an enhancement of star formation in
the regions affected by the outflow. However, both effects lead to a
reduction of the cold gas available, thus to a deep connection with
the overall star formation history of the host galaxy.

In order to investigate all these possible aspects of the AGN
phenomenon, a detailed description of the gas accretion onto SMBHs is
needed. Such a modelling should include all the different relevant
phases: (a) the loss of angular momentum $J$ in the gas component -
leading to gas infall towards the centre of the host galaxy, (b) the
accretion of this low-J material onto the SMBH and (c) the triggering
of the outflow and its feedback onto the host galaxy. The relevant
physical mechanisms involve a wide range of physical scales, ranging
from the {\rm Mpc}/{\rm kpc} scale characterizing the processes
destabilising the host galaxy, to the sub-pc scales of the accretion
region. Moreover, the dependence of the relevant mass/energy flows on
the physical properties of the host galaxy are still highly
uncertain. Inevitably, sub-grid models are still necessary in
hydro-simulations to define plausible physical dependencies and
explore the associated parameter space (see e.g. \citealt{Hopkins06f}
\citealt{Barai14}, \citealt{Sijacki15},
\citealt{Choi17}). Semi-analytic models (SAMs) have been mostly
adopted as flexible tools to probe the impact of AGN-driven winds
\citep{Fontanot06}, the origin of the downsizing trend
\citep{Hirschmann12}, the evolution of the AGN population along the
BH-Bulge relation \citep{Lamastra10}, and to explore the relation
between accretion rate efficiency and the spin of the SMBH
\citep{Fanidakis12}.

In this study, we present new implementations for BH accretion in the
state-of-the-art GAlaxy Evolution and Assembly ({\gaea}) semi-analytic
model. In our approach we assume that the cold gas present in galaxies
must lose most of its angular momentum $J$ before reaching the sub-pc
scales around the central SMBH, where it can be accreted and give rise
to the AGN phenomenon and its feedback. We model separately the
different timescales associated with the processes of $J$ loss and
SMBH accretion as well as the physics of AGN-driven outflows. The
inclusion of these timescales represents one of the main differences
with the standard modelling implemented in the previous versions of
the model, which assumed instantaneous accretion of the gas onto the
SMBH. For each relevant physical process, we consider different
options. In particular, we explore both empirically derived
prescriptions and fitting formulae suggested by numerical experiments
and analytic calculations.

This paper is organised as follows. In Section~\ref{sec:models} we
will describe our new modelling of SMBH accretion in {\gaea}. The
reference calibration set and a basic set of predictions will be shown
in Section~\ref{sec:results}, and we will discuss the implications of
our results in Section~\ref{sec:discussion}. Finally, we will
summarise our conclusions in Section~\ref{sec:final}.

\section{Semi-Analytic modelling}\label{sec:models}
In this paper we will define and compare different prescriptions for
BH growth in theoretical models of galaxy formation. In particular, we
will implement them in the GAlaxy Evolution and Assembly ({\gaea})
semi-analytic model (SAM). SAMs trace the evolution of galaxy
populations inside DMHs by modelling the main physical mechanisms
acting on the baryonic components, using physically and/or
observationally motivated prescriptions. These processes include
cooling and heating of baryonic gas, star formation, accretion of gas
onto SMBHs and the related feedback processes. The overall
architecture of these models results in a flexible tool to predict
galaxy properties for large galaxy samples, and allows a fast
exploration of the associated parameter space.

The {\gaea} model represents an evolution of the model published in
\citet{DeLuciaBlaizot07}. Key improvements with respect to the
original version include: (a) a detailed treatment of chemical
enrichment \citep{DeLucia14}, following explicitly the differential
enrichment associated with Asymptotic Giant Branch (AGB) stars, Type
II SNe and Type Ia SNe; (b) an updated modelling of stellar feedback
\citep{Hirschmann16}, featuring ejecting feedback in the form of
stellar-driven outflows (inspired by results from hydrodynamic
simulations), combined with a timescale of gas re-incorporation that
depends on DMH mass \citep{Henriques13}; (c) an improved modelling of
disc sizes \citep{Xie17}, that traces the evolution of angular
momentum following the mass and energy exchanges among different
galaxy components. In \citet{Fontanot17b}, we show that {\gaea} is
able to reproduce the evolution of the galaxy stellar mass function
and cosmic star formation rate up to the highest redshifts at which
measurements are available ($z\sim7$). We will refer to the reference
{\gaea} run published in \citet{Hirschmann16}, and based on the
``FIRE'' feedback scheme, as H16F. H16F is also able to reproduce the
gas fractions and mass metallicity relations at $z<3$, but it
overpredicts the activity levels of massive galaxies at low-z
\citep{Hirschmann16}, albeit correctly reproducing the fraction of
passive galaxies as a function of stellar mass and hierarchy
\citep{DeLucia19}. Moreover, the predicted size-mass and angular
momentum-mass relations for model galaxies (for both disc- and
bulge-dominated morphologies) are in relatively good agreement with
observational measurements both in the local Universe and at higher
redshift \citep{Zoldan18, Zoldan19}. Other versions of the {\gaea}
model include a treatment of the cold gas partition in atomic and
molecular hydrogen \citep{Xie17}, and a modelling for a variable
stellar initial mass function \citep{Fontanot17a, Fontanot18a}. These
two implementations will not be considered in the present study.
\begin{table}
  \caption{Parameter Calibration for the BH accretion models.}
  \label{tab:parameters}
  \renewcommand{\footnoterule}{}
  \centering
  \begin{tabular}{lcccc}
    \hline
    & H16F & {\ragaeaF} & {\ragaeaH} \\
    \hline
    \multicolumn{4}{c}{Stellar Feedback parameters}  \\
    \hline
    $\alpha_{\rm SF}$            & 0.03 & 0.1  & 0.1   \\
    $\epsilon_{\rm reheat}$       & 0.3  & 0.13  & 0.09  \\
    $\epsilon_{\rm eject}$        & 0.1  & 0.23  & 0.09  \\
    $\kappa_{\rm radio}/10^{-5}$   & 1.0  & 0.6  & 2.7   \\
    $\gamma_{\rm reinc}$          & 1.0  & 0.68  & 0.82 \\
    \hline
    \multicolumn{4}{c}{BH accretion parameters} \\    
    \hline
    $f_{\rm lowJ}/10^{-3}$         & --- & 6. & --- \\
    $f_{\rm BH}/10^{-3}$           & --- & 0.09  & --- \\
    \hline
    $R_0/{\rm kpc}$                    & --- & ---  & 2.0  \\
    $\eta_{\rm BH}$               & --- & ---  & 14.5 \\
    $\epsilon_{\rm qw}/100.$      & --- & 3.2 & --- \\ 
    \hline
    $f_{\rm cen}/10^{-3} $         & --- & 3.0  & 1.7 \\
    \hline
  \end{tabular}
\end{table}

The model for SMBH accretion used in H16F is the same as in
\citet{Croton06} and \citet{DeLuciaBlaizot07} and is based on
\citet[][KH00 hereafter]{Kauffmann00}. This implementation has known
shortcomings in reproducing the space density of luminous AGN at
high-redshift \citep{Marulli08}. In this model, SMBHs growth is
efficient only after mergers, where a fraction of cold gas is
instantaneously accreted onto the SMBH. This fraction depends on the
mass ratio ($m_{\rm rat}$) between the two merging galaxies, on the
total amount of available cold gas ($M_{\rm cold}$) and on the virial
velocity of the host DMH ($V_{\rm vir}$), and it is modulated by
the free parameter $f_k$:

\begin{equation}\label{eq:kh00}
\dot{M}_{\rm Q,KH00} = f_K \frac{m_{\rm rat} M_{\rm cold}}{ 1 + (V_{\rm vir} / 280 [km/s])^{-2}}.
\end{equation}

\noindent
This accretion channel gives rise to the most luminous AGN,
therefore it has been traditionally defined the QSO-mode of accretion.
Radio-mode accretion of hot gas is instead modelled following
\citet{Croton06}, and assumed to be proportional to the mass of the BH
($M_{\rm BH}$), to the virial velocity and to the fraction of the hot
gas in the DMH ($f_{\rm hot}$), modulated by the free parameter
$\kappa_{\rm radio}$:

\begin{equation}\label{eq:radio}
  \dot{M}_{\rm R,KH00} = \kappa_{\rm radio} \frac{M_{\rm BH}}{10^8 \msun}
  \frac{f_{\rm hot}}{0.1} \left( \frac{V_{\rm vir}}{200 [km/s]} \right)^3.
\end{equation}

\noindent
It is worth stressing that the model for the growth of SMBHs
implemented in H16F has been primarily calibrated against the
evolution of the galaxy stellar mass function (GSMF) at $z<3$. The
free parameters in Eq.~\ref{eq:kh00} and~\ref{eq:radio} have been
chosen to reproduce the local $M_{\rm BH}$-$M_{\rm bul}$ relation and
the evolution of the massive end of the galaxy stellar mass function
up to $z\sim3$. In our reference model, the merger channel is not
associated to an explicit AGN feedback on the cold gas component of
the host galaxy. In this work, we do not attempt to retune the H16F
run to reproduce the AGN-LF, and we just show the corresponding
predictions as a reference.

In this work, we will present new implementations to describe SMBH
accretion and its interaction with the host galaxy. We will mainly
focus on cold gas accretion, i.e. the QSO-mode, while leaving the
Radio-mode model as in Eq.~\ref{eq:radio}. Our improved treatment uses
a combination of empirical, numerical and analytic models, derived
from observational constraints, high-resolution controlled experiments
and analytic solutions to the equations governing the interplay
between the multiphase gas and the accreting SMBHs. We describe the
AGN phenomenon using a three step approach: (a) first, we model the
{\it $J$-loss} rate (Sec.~\ref{sec:inflow}) of cold gas in galactic
discs, responsible for its accumulation at the centre of the galaxy
and the creation of a gas {\it reservoir}\footnote{This gas reservoir
  mimics the role of the accretion disc around the SMBHs. We use the
  term reservoir following the same choice made in \citet{Fontanot06}
  in order to stress that in our modelling we do not attempt to
  resolve the detailed structure of the accretion disc around the
  central SMBH. Moreover, we assume that the small amount of material
  transferred from the gaseous disc does not modify its structure.}
around the SMBH; (b) we then predict the actual {\it accretion}
(Sec.~\ref{sec:accretion}) of cold gas from the reservoir onto the
central object; (c) finally, we estimate the gas removal from the
galaxy in AGN-driven {\it outflow} winds (Sec.~\ref{sec:outflow}). In
addition to these processes we also consider different alternatives
for the SMBH {\it seeding} (Sec.~\ref{sec:seeding}).

The calibration of the parameters we introduce in our new models has
been performed requiring them to reproduce the evolution of the AGN-LF
(Fig.~\ref{fig:agnlf}). The inclusion of a QSO-mode feedback in
{\gaea} implies a new and additional channel for gas heating and gas
ejection, and this perturbs the efficiency of stellar feedback in our
model. Therefore, we perform a retuning of the stellar feedback
parameters against the evolution of the GSMF. We check that this
approach is sufficient for our new {\gaea} runs to reproduce all
galaxy properties we discuss in previous papers (i.e. mass-metallicity
relations, quenched fractions, cold gas fractions; see
Appendix~\ref{app:b}).

We will consider {\gaea} predictions based on the merger trees
extracted from the Millennium Simulation \citep[][MS]{Springel05}, a
numerical realization of $500^3$ {\rm Mpc}$^{-3}$ cosmological volume
assuming the WMAP1 $\Lambda$CDM concordance cosmology
(i.e. $\Omega_\Lambda=0.75$, $\Omega_m=0.25$, $\Omega_b=0.045$, $n=1$,
$\sigma_8=0.9$, $H_0=73 \, {\rm km/s/Mpc}$). We do not expect the
mismatch of these parameters with respect to the latest constraints
\citep{Planck_cosmpar} to affect our main conclusions in a significant
way \citep[see e.g.][]{Wang08}. All stellar-based quantities are
computed assuming a universal Chabrier IMF. In order to test, the
effect of the merger trees resolution on our conclusions we also
consider predictions based on the Millennium-II Simulation
\citep[MSII][]{BoylanKolchin09}. The MSII span a smaller volume than
the MS on a grid of similar size, resulting in a 125 times better
particle mass resolution. We will discuss the comparison between MS
and MSII in the main text whenever appropriate, and we collect some
basic results in Appendix~\ref{app:resol}.

\subsection{BH seeding}\label{sec:seeding}
In the following, we adopt the same BH seeding scheme used in
\cite{Xie17}: every time a new DMH is resolved in the merger tree we
seed it with a BH mass ($M_{\rm BH}$) scaled with the parent DMH
mass\footnote{In the {\gaea} framework we use $M_{200}$ as a proxy for
  DMH mass.}  ($M_{\rm DM}$):

\begin{equation}
  M_{\rm BH} = \left( \frac{M_{\rm DM}}{10^{10} \msun h^{-1}} \right)^{1.33}
  \frac{10^{10} \msun h^{-1}}{3 \times 10^6 \msun},
  \label{eq:bhseed}
\end{equation}

\noindent
where the slope of the relation is derived from \citet{Volonteri11}
and~\citet{DiMatteo03}. Applying Eq.~\ref{eq:bhseed} to the $ M_{\rm
  DM}$ mass distribution, we get seed masses of the order of $\sim10^4
\msun$. We also test a ``fixed'' seeding scheme, by implanting the
same BH seed in each DMH. We test fixed seeds in the range $10^3-10^5
\msun$ and we find that the results discussed in the following
sections are mostly insensitive to the seeding scheme. A lower mass
seeding ($\lesssim 10^2 \msun$) is still able to reproduce lower
redshift constraints, but struggles in reproducing the luminous AGN at
high-redshifts. On the other hand, a flat seeding at $\gtrsim 10^5
\msun$ helps to reproduce the space densities at the bright-end of the
LF, but complicates the recovery of the faint-end of the LFs.

It is worth stressing that our seeding approach is rather
conservative. We do not consider the hypothesis that SMBHs can form
with masses larger than $\gtrsim 10^6 \msun$ via direct collapse of
giant gas clouds in the early Universe, as has been recently suggested
\citep[see e.g.][]{Valiante16, Pacucci17}, or from the accretion of
stellar mass BHs \citep{Boco20}. Assuming that a small fraction of DMHs in
our merger trees can host such initial SMBHs would help reconciling
predictions from our models with high-z observations. However, the
early phases of DMH assembly are poorly constrained at the MS
resolution that corresponds to an DMH mass resolution of
$10^{10}\msun$. Therefore we defer a more detailed study of the Early
Universe and the impact of seeding prescriptions on high-z sources to
future work.

\subsection{$J$-loss}\label{sec:inflow}
The first phase of the AGN triggering process requires that a fraction
of the cold gas available in the host galaxy loses enough angular
momentum to reach the central regions and become available for
accretion. In the following, we assume that the cold gas inflow
towards the central SMBH is triggered by both galaxy mergers and disc
instabilities. We model this phase by means of a gas reservoir that
mimics the accretion disc around the central SMBH, without any attempt
to model its detailed structure.

\subsubsection{SFR-driven $J$-loss}\label{sec:inflowA}
We consider two different approaches. The first one is an empirical
model developed in the framework of the semi-analytic code MOdel for
the Rise of GAlaxies aNd AGN ({\morgana} - \citealt{Monaco07}) and
first described in \citet[F06 hereafter]{Fontanot06}. In this model, a
fraction of the host cold gas is supposed to lose a substantial amount
of its angular momentum and accumulate in a gas reservoir of mass
$M_{\rm rsv}$ around the BH \citep[see also][]{Granato04}. This loss
of angular momentum is driven by physical mechanisms like turbulence
or radiation drag that are typically onset by SFR in the dense central
regions ($\psi_{\rm cen}$) of the host galaxy (i.e. a few {\rm kpc}
around the galaxy centre). In detail, we assume that the growth rate
of this reservoir is proportional to $\psi_{\rm cen}$, via a free
parameter $f_{\rm lowJ}$:

\begin{equation}\label{eq:f06jloss}
\dot{M}_{\rm J}^{\rm F06} = f_{\rm lowJ} \psi_{\rm cen}.
\end{equation}

\noindent
In galaxy mergers we assume that $\psi_{\rm cen}$ equals the amount of
SFR triggered by the merger itself and predicted by the collisional
starburst model \citep[$\psi_{\rm cs}$]{Somerville01}:

\begin{equation}
\psi_{\rm cen} = \psi_{\rm cs}.
\end{equation}

\noindent
Disc instabilities in {\gaea} do not affect the cold gas component in
model galaxies (i.e they do not trigger extra star formation), but
simply move a fraction of stars from the disc to the bulge of the host
galaxy to restore stability. However, disc instabilities are a
relevant channel for bulge growth at intermediate host galaxy masses
(i.e. $10^{10} < M_\star/\msun < 10^{11} $ \citealt{DeLucia11}) and
most models consider them as a viable AGN triggering channel (see
e.g. next section). Therefore, we define a $\dot{M}^{\rm DI}_{\rm
  F06}$ associated with discs instabilities and we assume it to be
proportional to the corresponding growth rate of the bulge component
$\dot{M}_{\rm bul}$:

\begin{equation}
\dot{M}^{\rm DI}_{\rm J} = f_{\rm lowJ} \mu \dot{M}_{\rm bul}
\end{equation}

\noindent
The constant $\mu$ is fixed by requiring that the amount of gas
inflowing in the reservoir during disc instabilities is similar to
that obtained for the scheme described in the next section. We check
that this condition is achieved for a $\mu=10$, so this is the value
that we adopt in this paper.

\subsubsection{$J$-loss following HQ11}\label{sec:inflowB}
As an alternative scenario, we consider the analytic approach proposed
by \citet[HQ11 hereafter]{HopkinsQuataert11}. Differently from the
previous prescription, in HQ11 angular momentum losses are not
modelled as a SFR-related mechanism, but as an effect of the overall
perturbations induced in the cold gas disc by instability events.

Every time the gas disc becomes unstable (either by a disc instability
or a merger), we assume that a fraction of its cold gas loses enough
angular momentum and becomes available for accretion onto the
SMBH. Following the analytic gravitational torque model proposed by
HQ11, we model the $J$-loss rate as:

\begin{equation}\label{eq:hq11jloss}
  \begin{array}{ll}
    \dot{M}_{\rm J}^{\rm QH11} & = \eta_{\rm BH} \left( \frac{R_0}{100 pc} \right)^{-3/2}
    \left( \frac{M_{\rm BH}}{10^8 \msun} \right)^{1/6} \times \\
     & \times f_d^{5/2} \left( \frac{M_{\rm disc}}{10^9 \msun} \right)
    \left( 1+\frac{f_0}{f_{\rm gas}} \right)^{-1}
  \end{array}
\end{equation}

\noindent
where $R_0$ represents the reference galactocentric distance used to
compute key quantities like:

\begin{equation}
  f_d =  M_{\rm disc}(R_0) / M_{\rm tot}(R_0)
\end{equation}

\begin{equation}
  f_{\rm gas} = M_{\rm gas}(R_0) / M_{\rm disc}(R_0)
\end{equation}

\begin{equation}
  f_0 = 0.31 f_d^2 \left( \frac{M_{\rm disc}(R_0)}{10^9 \msun} \right)^{-1/3}
\end{equation}

\noindent
In previous equations, $M_{\rm disc}$, $M_{\rm gas}$ and $M_{\rm tot}$
represent the disc, cold gas and total mass inside $R_0$. We compute
the latter using the scale radii for the disc, gas and bulge
components estimated by the model of \citet[see also
  \citealt{Zoldan17,Zoldan19}]{Xie17}.

\subsection{Accretion}\label{sec:accretion}
\begin{figure}
  \centerline{ \includegraphics[width=9cm]{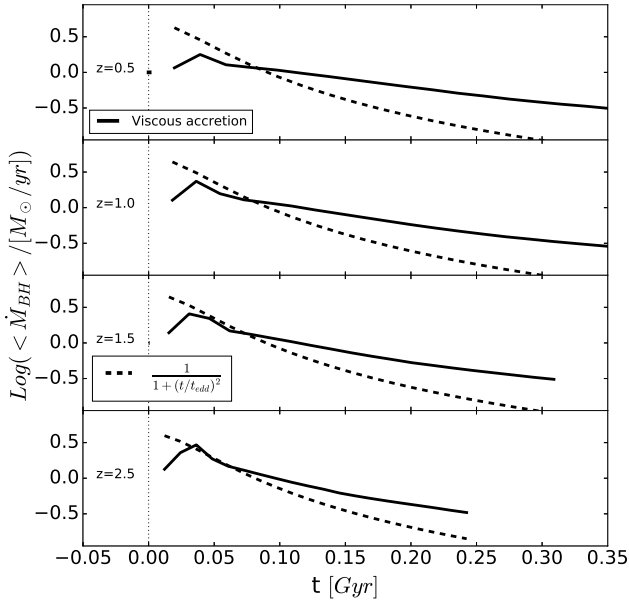} }
  \caption{Mean accretion rates onto a $M_{\rm BH} \sim 10^{8.5} \msun$
    in the {\ragaeaF} run (solid line) compared with the 
    power law decline assumed in the \citet{Hopkins06f} light
    curve. Both curves are normalized to the same total
    accretion. Different panels show predictions at different
    redshifts: the different dynamical range reflects the time
    interval between subsequent snapshots in the MS.}\label{fig:acr}
\end{figure}
\begin{figure*}
  \centerline{ \includegraphics[width=18cm]{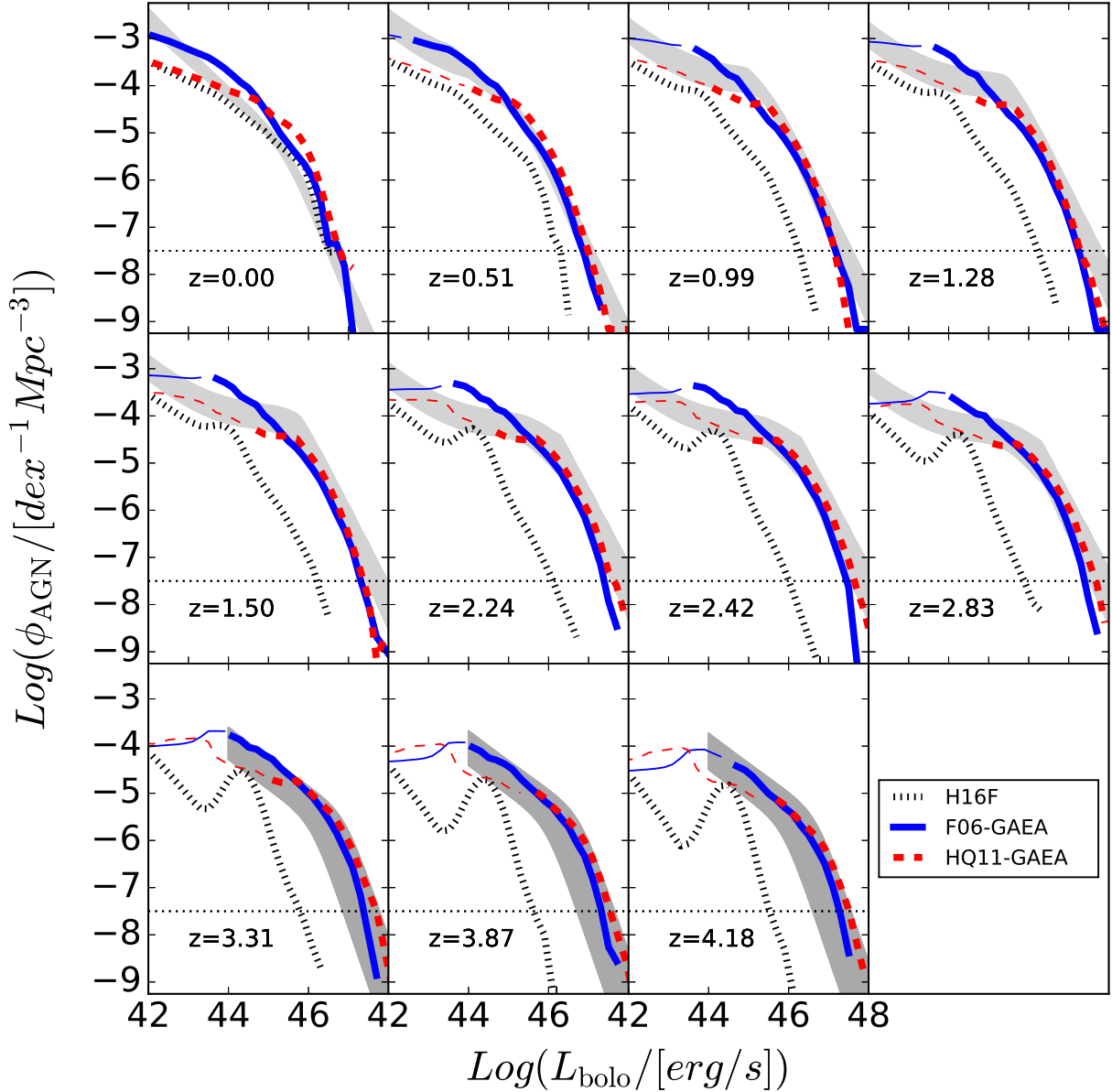} }
  \caption{Redshift Evolution of the AGN LF. Data are from
    \citet{Ueda14} and \citet{Fiore12}, with bolometric corrections as
    in \citet{Marconi04}. Black, red and blue lines represent
    predictions from the H16F (i.e. the \citealt{Kauffmann00} model),
    {\ragaeaH}, and {\ragaeaF} prescriptions respectively. Dotted
    lines represent the space density corresponding to 10 objects in
    the MS volume).}\label{fig:agnlf}
\end{figure*}
The material accumulated into the reservoir/accretion disc is
available for accretion onto the central SMBH. In this work, we do not
attempt a detailed modelling of the accretion disc evolution, but we
rely on two approximated prescriptions. We first test a prescription
explicitly accounting for the relation between the gas mass in the
reservoir and the mass of the central SMBH. We then test a more
general prescription applying a universal light curve to each shining
object: in this case we assume that the luminosity evolution traces
the accretion history.

The accretion prescriptions we consider, coupled with the amount of
cold gas available in the gas reservoir, lead to large accretion
rates. Therefore, we limit the accretion in a given time interval to a
rate:

\begin{equation}\label{eq:eddlim}
  \dot{M}_{\rm edd} = 100 \frac{L_{\rm edd}}{c^2} = 10 \frac{M_{\rm
      BH}}{t_{\rm edd}}
\end{equation}

\noindent
where $L_{\rm edd}$ represents the Eddington luminosity and $c$ the
speed of light. Please note that this definition correspond to ten
times the mass accretion rate for a SMBH with a radiative efficiency
of 10 percent, i.e. over an Eddington-Salpeter time scale $t_{\rm edd}
\sim 45 {\rm Myr}$ (as assumed in F06 and HQ11). This higher limit is
motivated by both observational and theoretical results \citep[see
  e.g.][]{Takeo19, Jiang19, Delvecchio20}. Theoretical models show
that such high accretion rates can be indeed frequent, especially at
high redshift \citep[see e.g.][]{Inayoshi16}.

\subsubsection{Viscous accretion}\label{sec:accrA}
As first choice, we consider an accretion rate determined by the
viscous accretion timescale. We assume the rate defined in
\citet{Granato04}:

\begin{equation}\label{eq:bhaf06}
  \dot{M}^{\rm F06}_{\rm BH} = f_{\rm BH} \frac{\sigma_B^3}{G} \left(
  \frac{M_{\rm rsv}}{M_{\rm BH}} \right)^{3/2} \left( 1+\frac{M_{\rm
      BH}}{M_{\rm rsv}} \right)^{1/2}
\end{equation}

\noindent
where $\sigma_B$ represents the velocity dispersion of the bulge.

\subsubsection{Light-curve model}\label{sec:accrB}
As a different scenario for gas accretion we consider the light curve
model proposed by \citet[see also \citealt{Hirschmann12}]{Hopkins06f}
and based on results from high-resolution hydrodynamical
simulations. Following each instability event, we assume that the AGN
light curve is characterized by two phases: a first regime where the
BH accretes exponentially at $\dot{M}_{\rm edd}$ until it reaches a
critical BH mass $M_{\rm BH}^{\rm crit}$:

\begin{equation}
  M_{\rm BH}^{\rm crit} = f_{\rm crit} 1.07 (M^{\rm in}_{\rm BH} + \Delta M_{\rm BH})
\end{equation}

\noindent
where $\Delta M_{\rm BH}$ represents the total mass accreted in the
event, $M^{\rm in}_{\rm BH}$ the initial mass of the SMBH and we fix
$f_{\rm crit} = 0.4$ as in \citet{Somerville08}.  Once the SMBH
reaches the critical mass, subsequent accretion is described as a
power-law decline as a function of the time ($\Delta t$) elapsed from
the peak accretion phase:

\begin{equation}\label{eq:hoplc}
   \dot{M}_{\rm BH}^{\rm QH11} = \frac{\dot{M}_{\rm edd}}{1+(\Delta t/t_{edd})^2}.
\end{equation}

\noindent
$\dot{M}_{\rm edd}$ represents the Eddington accretion rate of
the SMBH. In case of several triggering events close in
time, the clock is reset after each individual event.

\subsection{AGN-driven winds}\label{sec:outflow}
\begin{figure}
  \centerline{ \includegraphics[width=9cm]{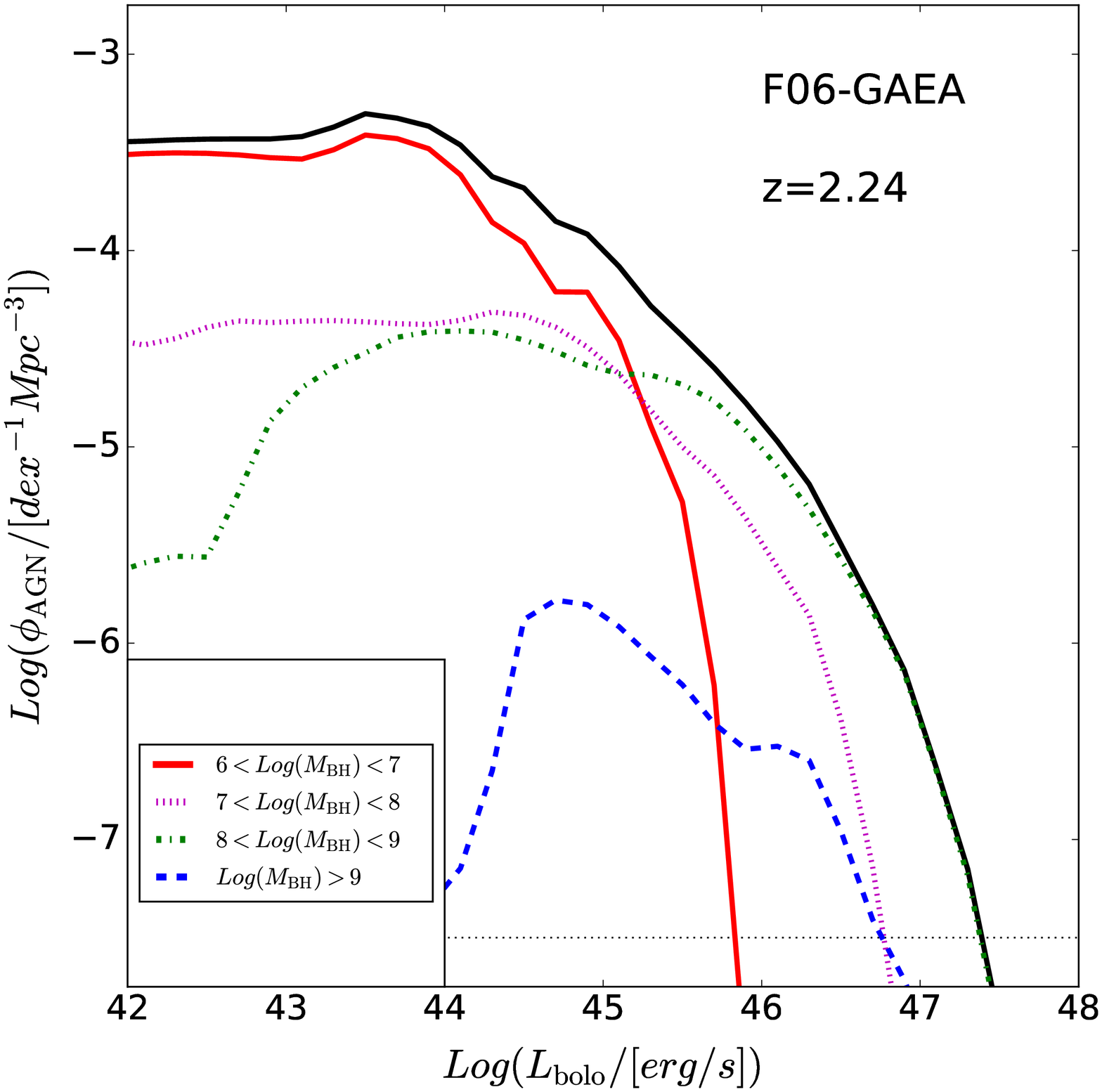} }
  \caption{AGN-LF at $z\sim2.25$ in the {\ragaeaH} run splitted into
    the contribution of SMBHs in different mass
    ranges.}\label{fig:splitlf}
\end{figure}
Finally, we consider the effect of AGN activity on the host galaxy,
and in particular on its cold gas phase. We assume that active AGN
exert a feedback effect of the surrounding medium, actively heating it
up and eventually expelling it in the hot gas in the form of a
AGN-driven galactic wind. In this study we will consider different
wind scenarios, either consistent with the results of analytic
calculations in idealized SMBH-galaxy disc geometries or based on
empirical arguments.

Moreover, following \citet{MonacoFontanot05}, we allow AGN-driven
winds to trigger further accretion onto the central SMBH. In this
scenario, AGN-driven winds are powered by the combined effect of SNe
explosions and radiation pressure of the AGN; we assume that a
fraction $f_{\rm cen}$ of the ISM is compressed to the centre, and
added to the BH reservoir. \citet{Mori02} suggest a value of the order
of $10^{-3}$ for $f_{\rm cen}$, assuming that $\sim$ 20 per cent of
the gas is compressed to the centre, but only $\sim$ 1 per cent of
this gas loses enough angular momentum to be eventually accreted onto
the reservoir. In the following, we treat $f_{\rm cen}$ as a free
parameter (see Table~\ref{tab:parameters} for the calibrated values).

\subsubsection{Empirical outflows}\label{sec:outflowA}
The last AGN phase we consider in our model is the outflow phase,
i.e. AGN-driven winds. We model this phase by assuming that each
accretion event in Eq.~\ref{eq:bhaf06} and~\ref{eq:hoplc} leads to an
outflow from the cold gas disc (AGN-driven wind), that is
characterized by a rate:

\begin{equation}
  \dot{M}^{\rm F17}_{\rm qw} = \epsilon_{\rm qw} \dot{M}_{\rm BH}
\end{equation}

\noindent
Such a scaling is consistent with predictions from hydrodynamical
cosmological simulations \citep{Brennan18} and with observational
constraints \citep[see e.g.][]{Fiore17}. $\epsilon_{\rm qw}$ is
treated as a free parameter of the order of $10^2$ according to
\citet{Brennan18} and~\citet{Fiore17}.

\subsubsection{M19 outflows}\label{sec:outflowB}
As an alternative approach, we also implement results of an analytic
model for AGN-driven outflows proposed by \citet[M19
  hereafter]{Menci19}. Their two-dimensional approach computes the
expansion of AGN-driven shocks in a galaxy disc with an exponential
gas profile, the outflow expansion in different directions with
respect to the plane of the disc, and the total mass outflow rate as a
function of global properties of the host galaxy and of the luminosity
of the central AGN. In this work, we take advantage of tabulated
numerical solutions for the fraction ($f_{\rm qw}$) of cold gas
ejected as a function of the bolometric luminosity $L_{\rm
  bol}$, the total gas mass $M_{\rm gas}$, and the virial velocity of
the parent DMH $V_{\rm vir}$:

\begin{equation}
  \dot{M}^{\rm M19}_{\rm qw} = f_{\rm qw}(L_{\rm bol}, M_{\rm gas}, V_{\rm vir})
\end{equation}

\subsection{Combining different prescriptions.}
In this paper, we consider two different prescriptions for each phase
connected to the AGN phenomenon, leading to several combinations of
possible realizations. In the following sections, we will discuss two
specific realizations that exemplify the influence of each individual
choice on model predictions. The first realization ({\ragaeaF}
hereafter - blue lines in all figures) closely resembles the scheme
implemented in the semi-analytic code {\sc morgana}
\citep{Fontanot06}: it uses the SFR-dependent $J$-loss
(Sec.~\ref{sec:inflowA}), the viscous timescale for gas accretion
(Sec.~\ref{sec:accrA}), and the outflow rate from the model developed
by M19 (Sec.~\ref{sec:outflowB}). This model loosely retains the
original description defined in F06 and uses two free parameters
$f_{\rm lowJ}$ and $f_{\rm BH}$. The values adopted for these two
parameters are consistent with those used in F06, although the two
models use different integration time intervals. $f_{\rm BH}=6 \times
10^{-3}$ is in good agreement with the estimated value in F06, while
our best estimate for $f_{\rm lowJ}=0.09 \times 10^{-3}$ is lower but
still consistent with F06.

The alternative realization ({\ragaeaH} hereafter - red lines in all
figures) uses the prescriptions calibrated on high-resolution
simulations by HQ11 (Sec.~\ref{sec:inflowB}), the light curve model
from \citet[\ref{sec:accrB}]{Hopkins06f} and the empirically
calibrated outflows (\ref{sec:outflowA}). This model includes three
free parameters: $R_0$ and $\eta_{\rm BH}$ plus the scaling of the
empirical wind model $\epsilon_{\rm qw}$. The original HQ11 work
tested several $R_0$ values (up to 100 {\rm pc}) and proposed
$\eta_{\rm BH} \sim 5$. In our model, we use $R_0=$2 {\rm kpc}, a
value that is larger than those tested in HQ11, but still
representative of the central region of model galaxies. As for
$\eta_{\rm BH}$, we use a value of $\sim$15, that is still consistent
with HQ11. It is important to remember that the original HQ11
framework has been defined on high-resolution hydrodynamical
simulations able to resolve the central regions of host galaxies on
sub {\rm kpc} scales. In {\gaea}, we assume that the mass
distributions of the relevant components follow smooth analytic
profiles (i.e. an exponential profile for both the gaseous and stellar
disc and a Jaffe profile for the bulge). Given the different
approaches for the structural modelling in the hydro-simulations and
in {\gaea}, the agreement between the values used for $R_0$ and
$\eta_{\rm BH}$ seems reasonable. Finally, $\epsilon_{\rm qw}=320$ is
consistent with the range of values allowed by the measurements show
in \citet{Fiore17}. A summary of the two runs with all the associated
parameters can be found in Table~\ref{tab:parameters}.

The two runs considered are representative of the range of results
span by all possible combinations of the 3 couples of prescriptions
described in previous sections (see Appendix~\ref{app:multimod}). In
particular, test runs exploring different combinations show that the
$J-$loss prescription, i.e. the total amount of cold gas available for
accretion, plays the most important role for reproducing the evolution
of the AGN population. The different accretion schemes we define in
Sec.~\ref{sec:accretion} provide similar evolution after calibration
is performed on the model predictions. Fig.~\ref{fig:acr} shows as a
solid line the mean accretion rates for $M_{\rm BH} \sim 10^{8.5}
\msun$ SMBHs at different redshifts in {\ragaeaF}. In the same figure,
the dashed lines represent the time evolution corresponding to the
functional form assumed in Eq.~\ref{eq:hoplc}, normalized to the same
total accretion (the dashed line shows only the power-law decline of
the light curve). The two accretion curves differ significantly: the
main difference lies in the viscous accretion predicting e-folding
times larger than $t_{\rm ed}$ by a factor of $\sim 5$. Nonetheless,
the overall trend of a power law decrease after an initial peak is the
same in the two approaches. These results imply that the number
density of intermediate luminosity AGN critically depends on the total
amount of low$-J$ cold gas available, more than on the detailed
description of accretion. Finally, the M19 model has been compared
with the empirical results from \citet{Fiore17}, showing that its
analytic results are in agreement with the observational
constraints. The two outflow models provide rather different
descriptions of the role of AGN activity in triggering galaxy wide
winds. We use the measurements by \citet{Fiore17} to define a purely
empirical model, and constrast it with a parameter free prescription
that implements the results of the analytic calculations presented in
M19. The two models assume correlated quantities as primary
dependencies to scale the mass loading factor of the wind, namely the
AGN bolometric luminosity and SMBH accretion rates. The main
difference between the two models lies in M19 taking into account also
the gravitational potential of the host galaxy and its gas content,
while the empirical prescription depends only on the central engine.

\section{Results}\label{sec:results}
\begin{figure}
  \centerline{ \includegraphics[width=9cm]{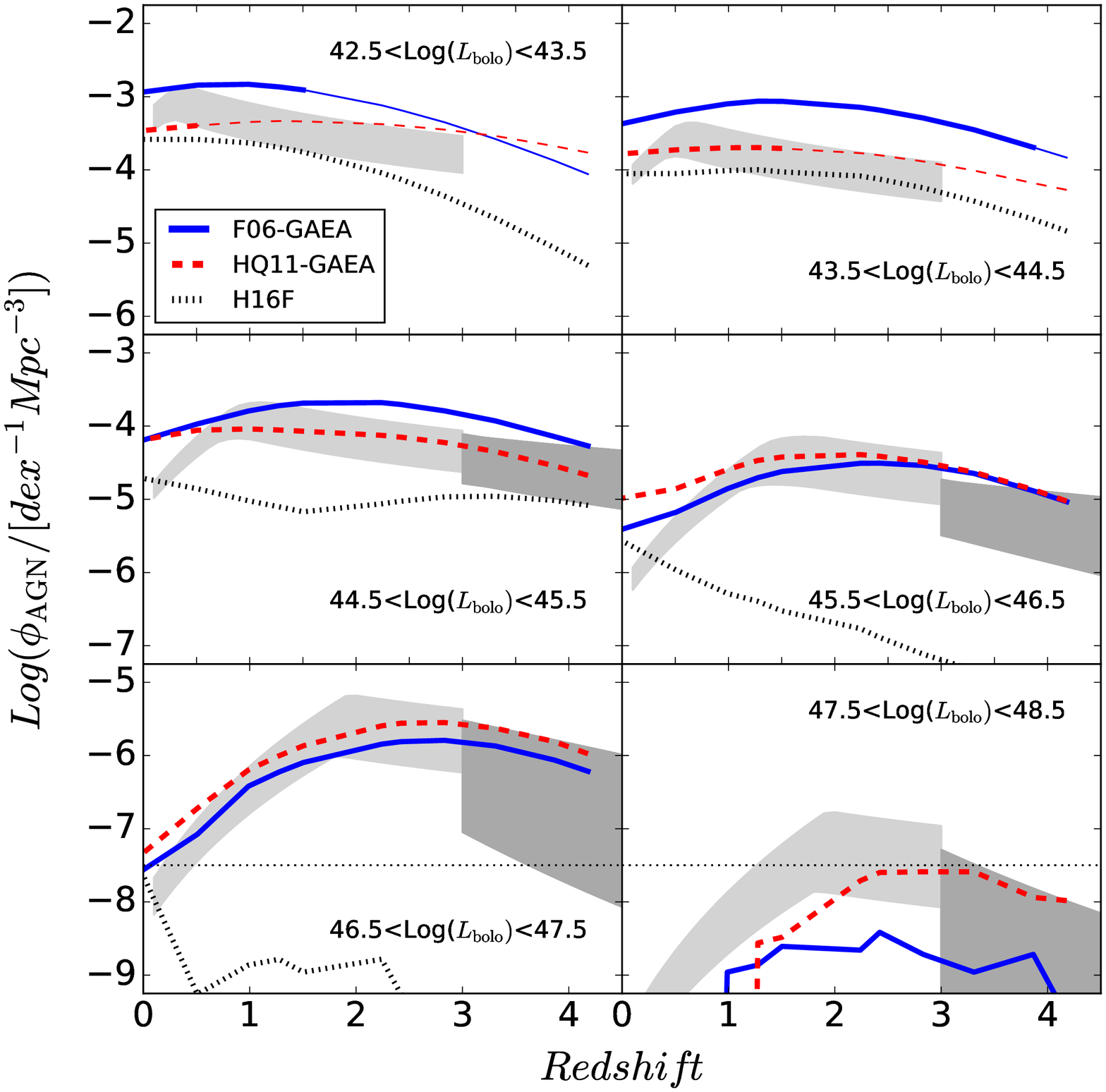} }
  \caption{AGN downsizing trends. Data are as in
    Fig.~\ref{fig:agnlf}. Black, red and blue lines refer to the
    different BH accretion models as in
    Fig.~\ref{fig:agnlf}.}\label{fig:ds}
\end{figure}
\begin{figure*}
  \centerline{ \includegraphics[width=9cm]{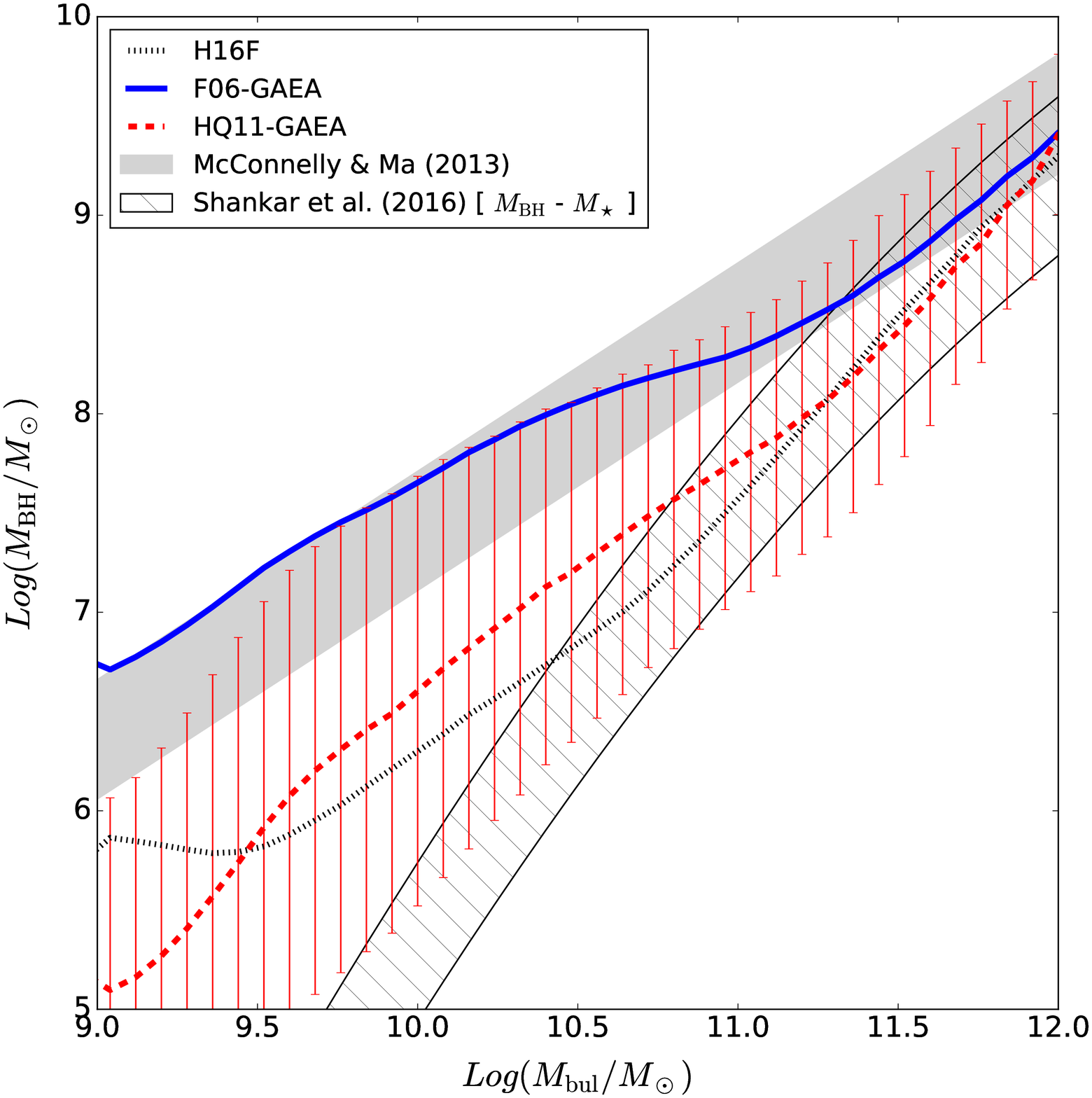}
    \includegraphics[width=9cm]{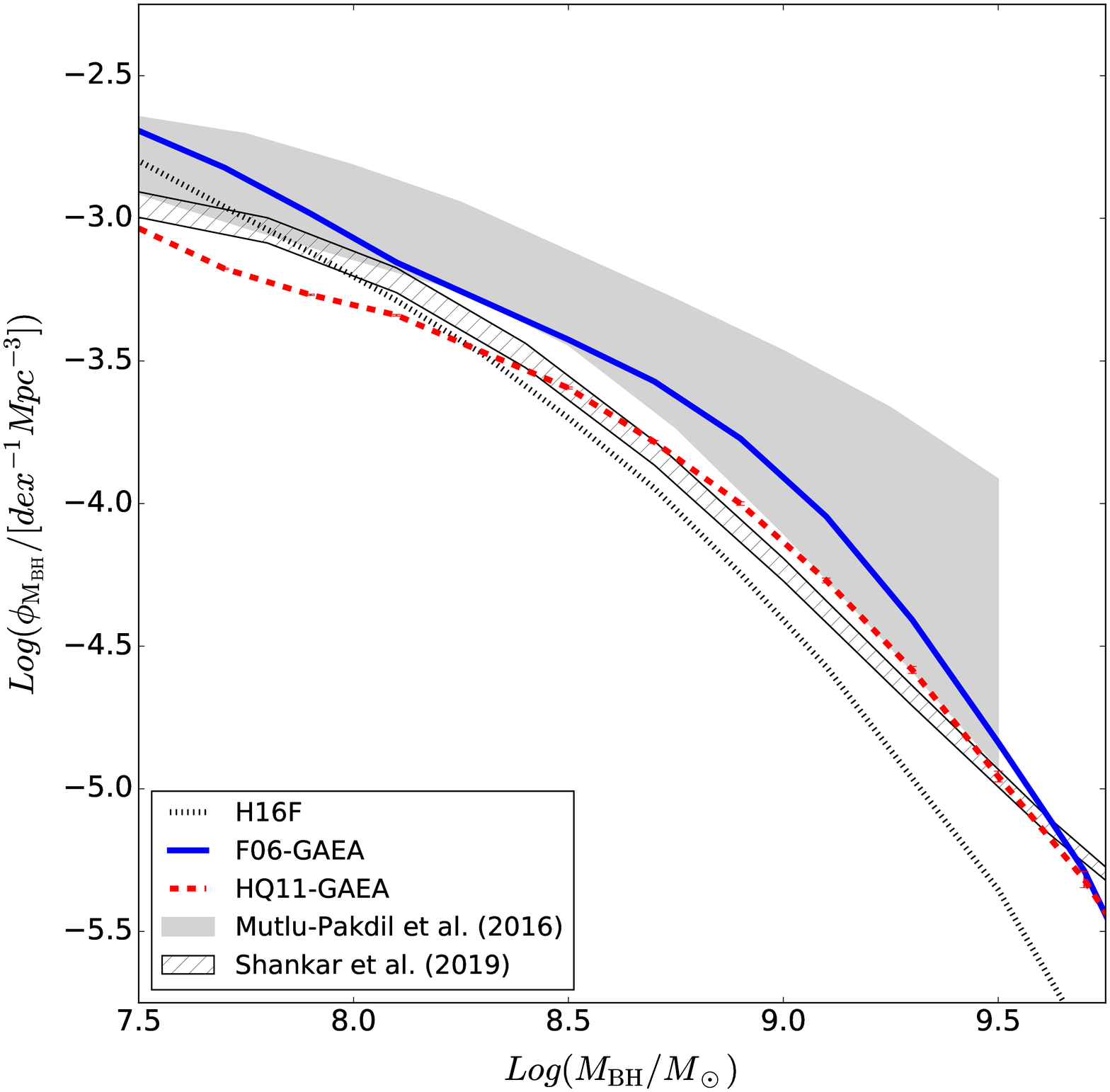} }
  \caption{{\it Left Panel}: $M_{\rm BH}$-$M_{\rm bul}$ relation at
    $z\sim0$. The shaded region shows the best-fit relations found by
    \citet{McConnellMa13}, respectively. The hatched area represents
    the intrinsic relation for the $M_{\rm BH}$-$M_\star$ relation
    proposed in \citealt{Shankar16}. {\it Right Panel}: z$\sim$0 BH
    mass function. Observational determinations are from a sample of
    local galaxies \citep[shaded area]{MutluPakdil16} or obtained
    using the ``accreted BHMF'' formalism \citep[hatched area see text
      for more details]{Shankar13b, Shankar19nat}. In both panels,
    black, red and blue lines refer to our {\gaea} realizations as in
    Fig.~\ref{fig:agnlf}.}\label{fig:bhrel}
\end{figure*}
In this section we collect a number of predictions on basic AGN
properties. In all panels we show the predictions from the original
recipe used in H16F with black lines, while blue ad
red lines are used for {\ragaeaF} and {\ragaeaH}. In most of the
following plots, the standard KH00 accretion model falls short in
reproducing the most luminous AGN (especially at $z>1$), in agreement
with \citet{Marulli08}, while our new models provide an overall better
description of the cosmic evolution of the space density of accreting
SMBHs.

In Fig.~\ref{fig:agnlf}, bolometric luminosities $L_{\rm bol}$ have
been obtained by assuming a radiative efficiency $\epsilon_{\rm rad}$
of 15 per cent for radiatively efficient AGN, i.e. with Eddington
ratios $f_{\rm edd}>0.1$

\begin{equation}\label{eq:accr_eff}
  L_{\rm bol} = \frac{\epsilon_{\rm rad}}{1-\epsilon_{\rm rad}} \dot{M}_{\rm BH} c^2 .
\end{equation}

\noindent
At lower Eddington ratios, we use results by \citet[see also
  \citealt{Hirschmann14}]{Churazov05}:

\begin{equation}
  L_{\rm bol} = 10 L_{\rm edd} f_{\rm edd}^2 .
\end{equation}

\noindent
Finally, sources with Radio-mode accretion (Eq.~\ref{eq:radio}) are
assumed to be very inefficient accretion systems, with a radiative
efficiency of 2 per cent, following results based on numerical
simulations (see e.g., \citealt{SadowskiGaspari17}). Model predictions
are compared with observational estimates for the AGN bolometric LFs.
At $z<3$, we start from the \citet[see also \citealt{Buchner15} for a
  different estimate of the LF]{Ueda14} hard X-ray LFs, while at
higher redshift we show measurements from \citet{Fiore12}. In both
cases, we apply bolometric corrections from \citet{Marconi04} to
estimate the expected bolometric LFs and we account in the calibration
for the uncertainties on the LF determination (represented by the grey
area in Fig.~\ref{fig:agnlf} and ~\ref{fig:ds}). We do not apply any
obscuration correction to model predictions and we compare the
intrinsic AGN-LFs to the observational constraints. This is motivated
by the choice of using the hard-X ray LF as primary constrain, and by
the choice of considering the full uncertainty range of the data in
the calibration procedure. Moreover, we check that the bolometric
AGN-LFs from \citet{Hopkins07b}, that is corrected for obscuration
effects, lie in the grey shaded area. We use the comparison between MS
and MSII predictions (Fig.~\ref{fig:appresol}) to estimate the
luminosity range where our models show a good level of convergence
(solid lines). For completeness, dashed lines show the predicted
AGN-LF below this luminosity limit. The convergence limit depends on
redshift, but also on the model considered, with {\ragaeaF} typically
reaching convergence at bolometric luminosities one order of magnitude
fainter than in {\ragaeaH}. This result is likely connected to the
modelling of the $J-$loss in the HQ11 framework. This approach heavily
relies on a proper description of the mass distribution between the
different galaxy components (i.e. disc and bulge), on their relative
profiles and scale radii. \citet{Xie17} provides a relevant
improvement in the {\gaea} framework, nonetheless, such a description
still represents one of the strongest limitations of our SAM, in
particular for low-mass objects and at lower resolutions.

The luminosity function based on the H16F run shows a peculiar shape,
in particular at high-z. At the faint end, the power law tail is
dominated by sources in the Radio-mode phase (Eq.~\ref{eq:radio}),
while the almost-Gaussian peak at bright luminosities is populated by
model galaxies in the merger-driven accretion phase
(Eq.~\ref{eq:kh00}). In particular, the narrow width of the latter
feature reflects the assumption of instantaneous accretion during
mergers. We stress that the original H16F run has not been explicitly
calibrated to reproduce the evolution of the AGN-LF in this
paper. Previous work by \citet{Marulli08} showed that the KH00 scheme
is not able to reproduce the evolution of the AGN-LF above $z>1$, in
the framework of the \citet{DeLuciaBlaizot07}. On the other hand, both
{\ragaeaF} and~{\ragaeaH} provide consistent predictions for the
AGN-LF up to $z\sim3$. At the bright end of the LF, the two models
provide similar predictions, in agreement with the observational
constraints up to the space densities robustly sampled by the MS (the
dotted lines in Fig.~\ref{fig:agnlf} mark the space density
corresponding to 10 objects in the MS volume). At fainter magnitudes,
the differences between the two models are more significant, with
{\ragaeaF} predicting a systematically larger number of sources below
the knee of the LFs (although number densities are still roughly
consistent with observational constraints). As we mention in previous
section, we checked that the differences between the two runs are
mostly due to the $J$-loss prescription, with the accretion
prescription playing a minor role. Indeed, the SFR-driven $J$-loss
predicts that more cold gas gets destabilized and goes into the
reservoir with respect to the HQ11 scenario. The LFs predicted by both
models are also characterized by several features. These features are
mainly due to the relative contribution of AGN powered by central
SMBHs of different $M_{\rm BH}$, as highlighted in
Fig.~\ref{fig:splitlf} (for clarity we show results for a
representative redshift range, but similar results hold at all cosmic
epochs). This figure shows the relative contribution to the LF of
SMBHs of different mass, and helps breaking the degeneracy between
$M_{\rm BH}$ and $\dot{M}_{\rm BH}$ in determining the AGN luminosity.

Another critical tests for our accretion models is the so-called AGN
downsizing, i.e. the evidence that the space density evolution of AGN
peaks at a decreasing redshift with increasing luminosity
(Fig.~\ref{fig:ds}, where we show only luminosity bins above the
convergence limit). This is just an alternative way to show the
evolution of the LF with respect to Fig.~\ref{fig:agnlf}. Our models
reproduce relatively well the observed trends between $10^{43}$ erg/s
and $10^{47}$ erg/s. At higher luminosities, the number of objects is
quite low; given the volume of the MS we cannot conclude if there is a
problem for our accretion schemes or this regime is affected by sparse
sampling effects (dotted lines in Fig.~\ref{fig:ds} represent the
space density corresponding to 10 objects in the MS volume).
%

We then consider some standard predictions for the mass distribution
of the $z=0$ SMBH population, i.e. its mass function (BHMF) and the
BH-Bulge relation. We recall that we perform the calibration of the
relevant parameters for our realizations on the AGN-LF and its
redshift evolution, therefore additional quantities represent real
predictions for our models. Other models in the literature use the
local BH-Bulge relations and/or the BHMF as the main calibration
set. We choose to focus on the AGN-LF for two main reasons. First of
all, a calibration on the local relation does not guarantee to
reproduce the evolution of the AGN-LFs, that represent a stronger
constraint for the differential SMBH accretion history. Moreover, the
observed slope and normalization of the BH-Bulge relations may be
seriously affected by selection biases \citep{Shankar16}. A different
normalization of the BH-Bulge relation also impacts the BHMF, through
a different estimate for the radiative efficiency.

In the left panel of Fig.~\ref{fig:bhrel} we show the BH-Bulge
relations in our realizations: these predictions agree within
1-$\sigma$ with the \citet{McConnellMa13} (or \citealt{KormendyHo13})
relation (grey shaded area). The typical scatter in the model
predictions
%
%
is sensibly larger than in the observational estimates. Although the
{\ragaeaH} run systematically underpredicts $M_{\rm BH}$ at fixed
$M_{\rm bul}$ with respect to the \citet{McConnellMa13}, these results
would still be consistent with the observed data, if we consider
selection biases as proposed in \citet{Shankar16}. The intrinsic
relation between $M_{\rm BH}$ and the the {\it total} stellar mass of
the host galaxy $M_\star$ is shown for reference in the left panel of
Fig.~\ref{fig:bhrel} as an hatched area. \citet{Shankar19nat} show
that a model consistent with the AGN-LF and assuming $\epsilon_{\rm
  rad}>0.1$ naturally falls up to one order of magnitude below the
observed relation for $M_{\rm BH} \lesssim 10^{10.5} \msun$. This
effect seems at play in {\ragaeaH}, while the {\ragaeaF} predictions
are aligned with the observed relation. {\ragaeaF} overpredicts the
space density of $L_{\rm bolo} \lesssim 10^{45} erg/s$ with respect to
{\ragaeaH} and observed data, at all redshifts
(Fig.~\ref{fig:ds}). Although the predicted luminosity functions are
still consistent with the observed AGN-LFs, this effect implies larger
accretion on (and growth of) $M_{\rm BH} \sim 10^6 - 10^7
\msun$. Finally, the steepness of the BH-Bulge relation obtained from
the KH00 prescription has been previously reported (see
e.g. \citealt{Arora19}).

Both runs predict similar trends for the z$\sim$0 BHMF
(Fig.~\ref{fig:bhrel} - right panel). In detail,the {\ragaeaH} run
predicts a lower space density of low-mass SMBH, with respect to
{\ragaeaF} and H16F, while both our new models predict a larger space
density of massive SMBHs than in H16F. As a reference we compare these
predictions with observational constraints derived either from a
sample of local galaxies \citep{MutluPakdil16} or from the AGN-LF
using the ``accreted/relic BHMF'' formalism and assuming a radiative
efficiency of 15 percent \citep{Shankar19nat}. We note that the
accreted BHMF has been computed via the continuity equation formalisms
developed by \citet{Shankar13b}. We have neglected mergers and
assumed, for simplicity, a constant Gaussian Eddington ratio
distribution peaked at -0.6 and with a width of 0.4 dex. The shape of
the resulting BHMF is not very sensitive to the shape of the input
Eddington ratio distribution, at least for $M_{\rm BH} > 10^8 \msun$
(see discussions in \citealt{Shankar13b}). We find a reasonable
agreement between the predictions of our models and the available
constraints.

\subsection{Effect of AGN-driven winds of the SFR of host galaxies}
\begin{figure}
  \centerline{ \includegraphics[width=9cm]{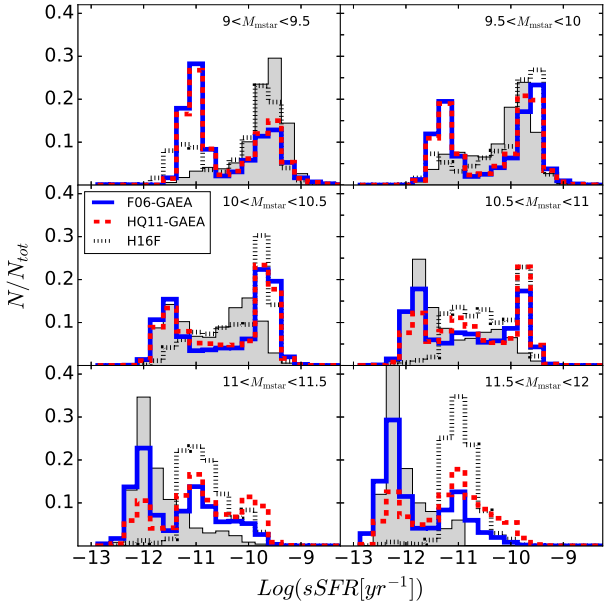} }
  \caption{sSFR distributions for model galaxies. Grey histograms
    represents the distribution of SDSS-DR7 galaxies with
    $0.025<z<0.05$. In all panels, black, red and blue lines
    refer to the different BH accretion models as in
    Fig.~\ref{fig:agnlf}.}\label{fig:sfrhisto}
\end{figure}
\begin{figure*}
  \centerline{ \includegraphics[width=9cm]{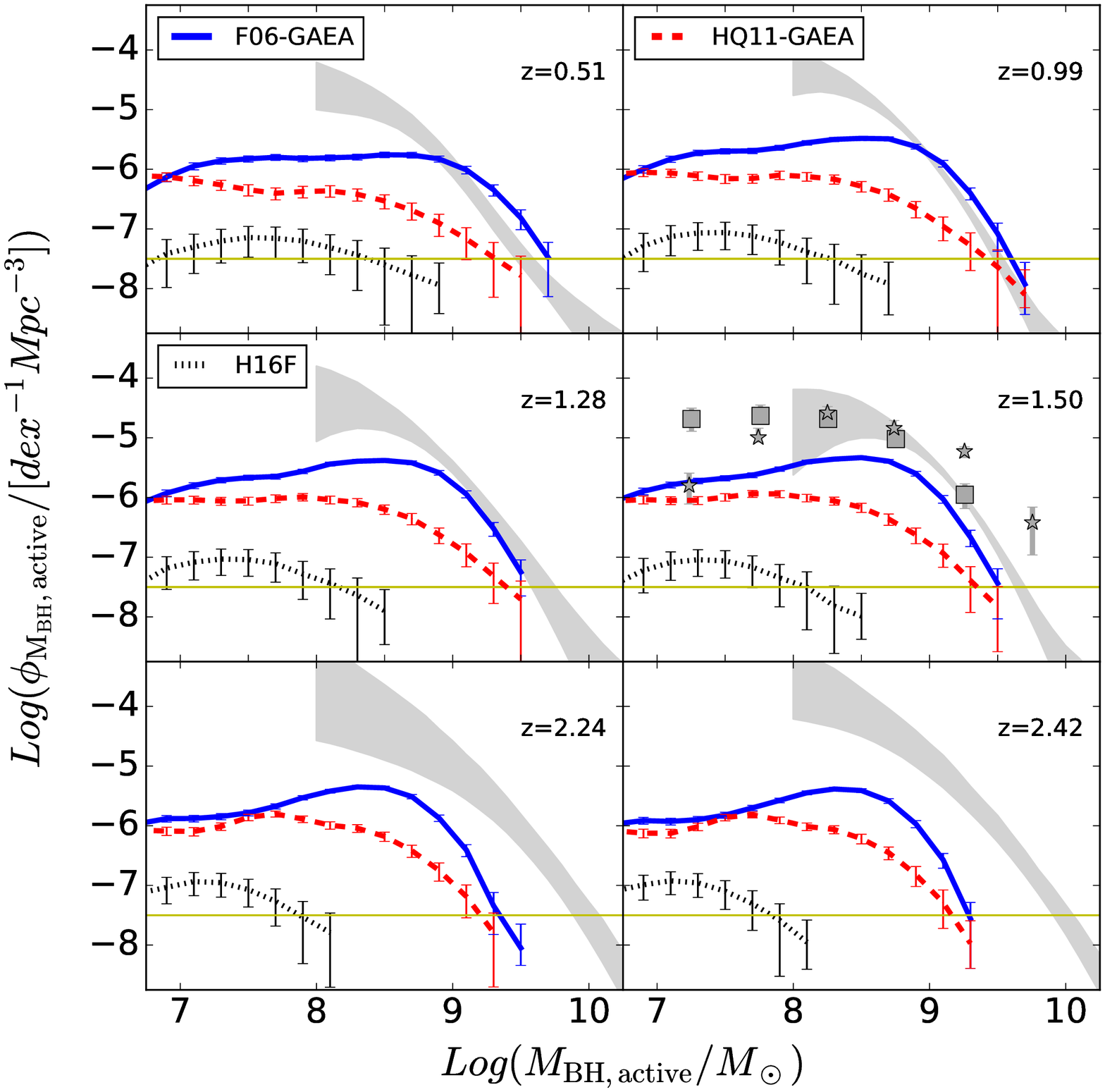}
    \includegraphics[width=9cm]{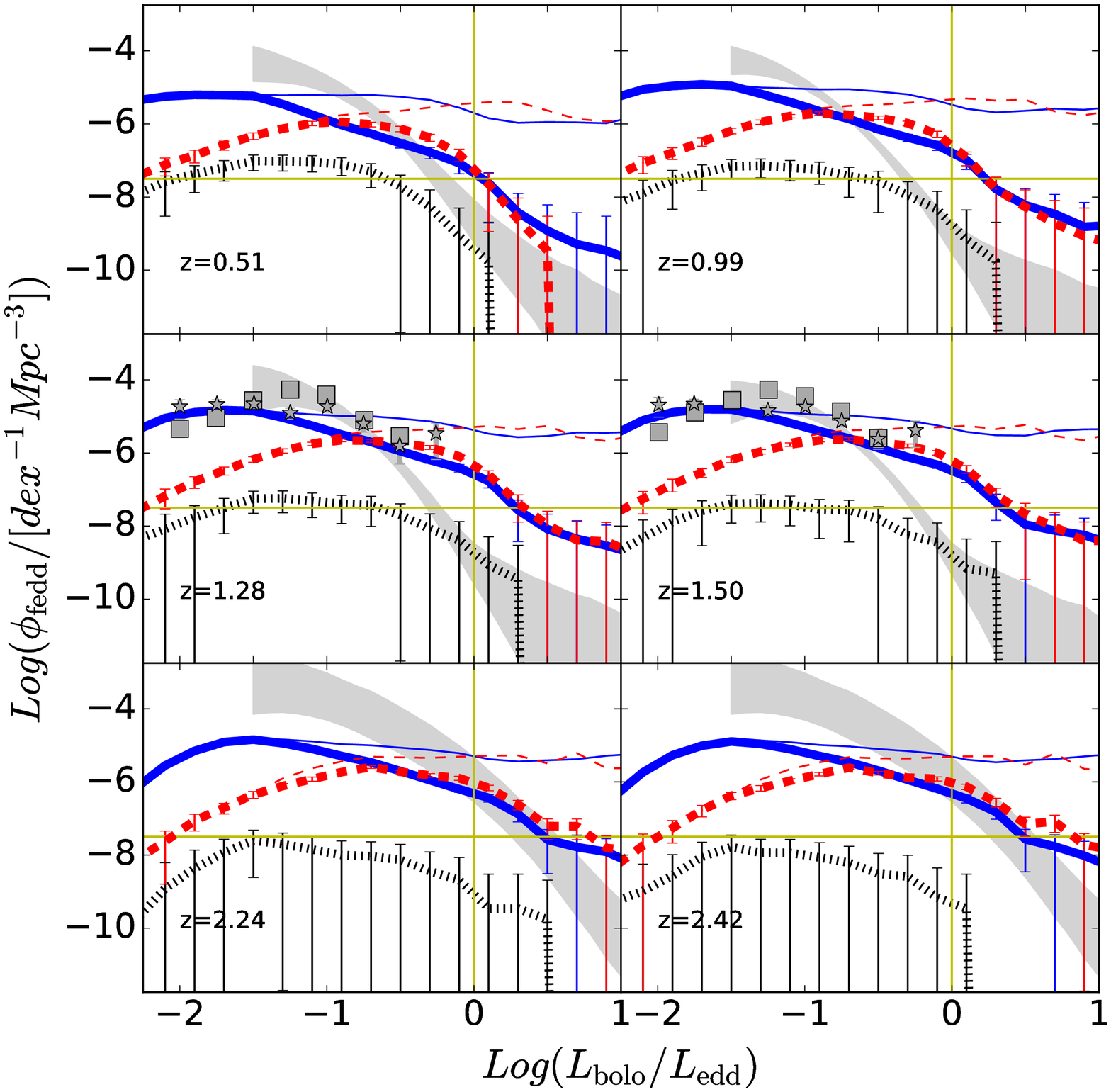}
  }
  \caption{{\it Left panel}: Redshift evolution of the mass function
    of {\it active} BHs, defined as sources powering AGN more luminous
    than $L_{\rm bol}>10^{44.6} erg/s$. {\it Right panel}: Redshift
    evolution of the probability distribution function of Eddington
    Ratios. Black, red and blue lines correspond to {\gaea} runs as in
    Fig.~\ref{fig:agnlf}. Dashed lines refer to the sample of model
    AGNs more luminous than $L_{\rm bol}>10^{44.6} erg/s$, while
    solid lines show the distribution for a sample selected using both
    the bolometric luminosity ($L_{\rm bol}>10^{44.6} erg/s$) and BH
    mass ($M_{\rm BH} > 10^8 \msun$) selections. Data are from
    \citet[shaded area]{KellyShen13} and~\citet[dots with
      errorbars]{Schulze15}.}\label{fig:eddlf}
\end{figure*}

The overall mass assembly history of galaxies in our runs implementing
a different BH accretion history is consistent with the results
presented in H16F (See Appendix~\ref{app:b}). The main difference with
respect to H16F lies in the predicted SFRs. We recall that this
observable represents a significant tension between the reference
{\gaea} model and observational data (see Fig.~8 in H16F). In the
standard {\gaea} run, galaxies with $M_\star > 10^{11} \msun$ at
z$\sim$0 have residual specific SFRs (sSFR$=$SFR/$M_\star$) larger
than those observed in the local Universe. Moreover, the distribution
of galaxies at intermediate masses ($M_\star \sim 10^{10} - 10^{11}
\msun$) does not show the observed bimodality. It is important to keep
in mind that the SFR levels predicted for the more massive galaxies
are low enough that most of these galaxies would still be classified
as passive \citep{DeLucia19}.

All {\gaea} runs we consider implement Radio-mode feedback in the form
of Eq.~\ref{eq:radio}. We have verified that this SFR excess for
massive galaxies cannot be solved by a different (or stronger)
implementation of Radio-mode feedback (see also
\citealt{Hirschmann16}), as it is not related to late cooling flows in
massive DMHs, nor to cold gas brought in by satellite galaxies. The
main reason for this residual SFR is the large cold gas content in the
main progenitors of these galaxies at $z\sim2$, and the low assumed
SFR efficiency, leading to low gas consumption rates. AGN-driven winds
represent an additional mechanism to remove cold gas from galaxies,
even from isolated ones. Therefore, they may represent a viable
solution to this problem. In Fig.\ref{fig:sfrhisto}, we compare the
sSFR distribution for galaxies in different mass bins with data from
SDSS galaxies (shaded distributions, in the $0.025<z<0.05$ redshift
range, where we expect the sample to be volume complete down to $10^9
\msun$). We assign an observationally equivalent SFR upper limit to
model galaxies with SFR $<10^{-4} \msunyr$ using the relation:

\begin{equation}
log(SFR) = 0.5*Log(M_\star)-6.59
\end{equation}

\noindent
This reproduces the locus of the upper limits of passive galaxies in
the SDSS sample. We perturb this relation adding a lognormal scatter
of 0.25 dex to the final SFR. When compared to the H16F predictions,
both our new realizations provide an improvement for the $M_\star >
10^{11} \msun$ mass bins. The improvement is more relevant for
{\ragaeaF}, while is relatively smaller for {\ragaeaH}: the number of
(almost) completely quenched galaxies is increased, but the shape of
the distribution still sensibly differs from observations. This
relative improvement also comes at the price of a clear overprediction
of the passive population in the $M_\star < 10^{10} \msun$ mass bins,
that breaks the agreement found in H16F, in this mass range. At
intermediate masses, {\ragaeaF} provides a marginal improvement in
recovering the predicted bimodality of the sSFR distributions.

The differences between the new runs and H16F are due to the
combination of different effects. The implementation of AGN-driven
winds plays a relevant role, by removing cold gas from galactic
discs. Nonetheless, it is worth stressing that the differences in the
quenching efficiencies between {\ragaeaF} and {\ragaeaH} depend mainly
on the $J$-loss prescription, i.e. on the total amount of cold gas
available for accretion onto SMBHs. In particular, the fraction of
quenched $M_\star>10^{11} \msun$ galaxies is always larger in
{\ragaeaF} than in {\ragaeaH}, irrespective of the outflow model
implemented. A second competing effect in regulating the SFR levels in
model galaxies is connected to the different stellar feedback
parameters involved in the calibration of {\ragaeaF} and {\ragaeaH}
runs. Both models assume an $\alpha_{\rm SFR}$ parameter three times
larger than H16F: this implies that the more efficient cold gas
depletion in massive galaxies is due both to the AGN-driven winds and
to the higher star formation efficiency. It is worth stressing that
despite the different stellar feedback parameters (in particular the
$\epsilon_{\rm reheat}$ parameter is roughly three times smaller than
H16F) the agreement of our models with the observed galaxy stellar
mass functions and stellar/gas mass-metallicity relations is similar
to that shown in H16F (see Fig.~\ref{fig:appevo}). These results
clearly show that the implementation of a different AGN modelling does
not heavily impact on the self-regulation of star formation and
feedback in galaxies.

The sSFR distribution predicted by our models is affected by the
modelling of AGN-driven winds. In particular, since the M19
prescription does not use free parameters, the red dashed histograms
are a direct prediction for the {\ragaeaH} run. On the other hand, in
the empirical prescription from \citet{Fiore17} the efficiency of the
mass-loading factor $\epsilon_{\rm qw}$ is treated as a free parameter
that can be varied to try to improve the match with the sSFR
distribution. The blue histograms in Fig.~\ref{fig:sfrhisto} show the
difficulties that also for the {\ragaeaF} run we face difficulties in
reproducing the SFR levels observed over a wide range of galaxies
stellar masses. It is possible to improve the agreement with the
observed distributions at the high-mass end of the GSMF, by increasing
the strength of cold gas removal in AGN-driven winds. This, however,
comes at the price of increasing the fraction of quenched galaxies at
the low-mass end of the GSMF, exacerbating the discrepancy with the
data, over this galaxy mass range. Our results are not in tension with
apparently opposite conclusions reached by, e.g., hydro-dynamical
simulations \citep{Brennan18}. Indeed, the tensions we see are mainly
due to our attempt to reproduce {\it at the same time} the evolution
of the GSMF, the AGN-LF, the mass-metallicity relations and the sSFR
distribution.

Overall, we conclude that AGN-driven winds alone are not a viable
solution to improve the agreement between our predicted sSFR
distributions and observational measurements. Additional modifications
of the stellar feedback prescription (taking into account also the
coupling between stellar and AGN-driven winds - see
e.g. \citealt{MonacoFontanot05}) have likely to be taken into account
in order to reproduce the detailed distribution of SFR of galaxies of
different mass, and will be the subject of future work.

\subsection{Eddington Ratios Distributions}\label{sec:eddlf}
In order to understand if our runs can capture the complexity of BH
accretion events at different cosmic epochs, we also consider the
evolution of the mass function of {\it active} BHs (aBHMF) and the
probability distribution functions for the Eddington ratios predicted
by our runs. In Fig.~\ref{fig:eddlf}, we compare these quantities with
observational estimates from \citet{KellyShen13}
and~\citet{Schulze15}. It is worth stressing that the BH mass estimate
is a difficult measurement, typically done via analysis of the MgII
and/or H$_\beta$ feature in QSO spectra. These measurements are
reliable\footnote{As an example, the contamination from the FeII line
  \citep[e.g.][]{Tsuzuki06} is an important source of uncertainty for
  MgII measurements \citep{Bischetti17, Vietri18}.} only for a
subsample of bright QSOs and for relatively massive SMBHs ($M_{\rm BH}
\gtrsim 10^7-10^8 \msun$). In the following, we use as reference the
results from the sample of \citet[shaded area]{KellyShen13}, that is
defined using Type I QSOs brighter than $L_{\rm bol} > 10^{44.6}
erg/s$. We also consider data from \citet[symbols with
  errorbars]{Schulze15}. These are in general consistent with the
\citet{KellyShen13} sample, but cover a smaller redshift range. We
apply to our model predictions the same luminosity cut as
\citet{KellyShen13} and we also apply a $M_{\rm BH} \gtrsim 10^8
\msun$ cut to account for the limitations due to SMBH mass
estimates. In order to account for the Type I selection, we simply
assume that these sources account for 25\% of the total population at
all redshifts and luminosities. Given the uncertainties in the
observational determinations (that require an extrapolation from a
small initial sample), the agreement of our predictions with the data
(Fig.~\ref{fig:eddlf} - left panel) is encouraging. In particular,
{\ragaeaF} is consistent with the \citet{KellyShen13} results within
$<1$ dex up to $z \sim 1.5$, while {\ragaeaH} is systematically below
the observed aBHMF by more than 1 dex. The differences between
{\ragaeaF} and {\ragaeaH} can be ascribed to the different AGN space
densities around the knee of the LFs, and to the different modelling
of $J$-loss and accretion. In particular, the larger space densities
of active SMBHs in {\ragaeaF} are due to the larger population of BHs
entering the adopted luminosity cut, and to the larger gas reservoirs
around the central SMBHs (the latter provide longer timescales for
accretion). In detail, we check that the $J$-loss prescription is more
important than the accretion prescription in determining the
aBHMF. There is a clear deficit of active SMBHs, in both models, at
$z>2$ and for $M_{\rm BH}>10^9 \msun$. This could be due to an
intrinsic problem for our modelling and/or to the conservative seeding
approach (sec.~\ref{sec:seeding}), that is not able to catch the early
stages of SMBH assembly via direct collapse of massive gas cloud.

An important point to keep in mind for further discussion is that the
predicted aBHMF extends well below $M_{\rm BH} \sim 10^8 \msun$,
which represents the confidence region for the observational
datasets. This effect has important consequences on the interpretation
of the predicted probability distribution function of Eddington
ratios. In the right panel of Fig.~\ref{fig:eddlf} we compare the
estimates from \citet{KellyShen13} with two different samples of model
AGNs. Solid lines correspond to an AGN sample selected applying to our
models both a cut in luminosity ($L_{\rm bol}>10^{44.6} erg/s$) and BH
mass ($M_{\rm BH}>10^8 \msun$): this sample shows a reasonable
agreement with the observational constraints (especially considering
that we are at space densities at the limit of MS resolution - the
dotted line marks the space density corresponding to 10 objects in the
MS). The situation changes dramatically if we include in the
distribution sources powered by smaller BHs (i.e. if we only consider
a luminosity cut). The resulting probability distribution functions
are shown as dashed lines: they are flat over a wide range of
Eddington ratios, clearly overpredicting the estimated space density
at high Eddington ratios.

These conclusions are consistent with the analysis of the contribution
to the AGN-LF of AGN powered by SMBHs of different mass
(Fig.~\ref{fig:splitlf}): while QSOs on the bright-end of the LF are
powered by SMBHs more massive than $\sim 10^8 \msun$, objects in the
mass range $10^6-10^7 \msun$ account for roughly half of the sources
below the knee of the LFs. It is currently quite challenging testing
this prediction of our models via current facilities, but this mass
range should be accessible with the next generation of space and
ground instruments (like Athena and JWST).

Our results suggest that massive SMBHs are in a ``self-regulated''
regime, i.e. the systems are able to auto-regulate the amount of cold
gas that is available for accretion. Lower mass BHs have not yet
reached this regime, and live in environments where large amounts of
gas is available for accretion. The flat distribution of Eddington
ratios is a natural consequence of our simplified assumption of a flat
limiting Eddington accretion rate (Eq.~\ref{eq:eddlim}) coupled with a
fixed luminosity cut.

It order to overcome this limit of our model, i.e. predicting a
probability distribution function of Eddington ratios that reproduces
the \citet{KellyShen13} estimates using only a luminosity cut,
relevant improvements in our schemes are needed. In particular,
preliminary work shows that an evolution of the limiting Eddington
rate as a function of $M_{\rm BH}$ is required in order to reduce the
contribution of $M_{\rm BH}<10^7 \msun$ to the LF at intermediate
bolometric luminosities. However, how this effect could be achieved in
our framework in a physical way is beyond the aim of the present
study. We plan to deepen this point by exploring alternative accreting
schemes and feedback scenarios in a future work.

\section{Discussion}\label{sec:discussion}
\begin{figure*}
  \centerline{ \includegraphics[width=18cm]{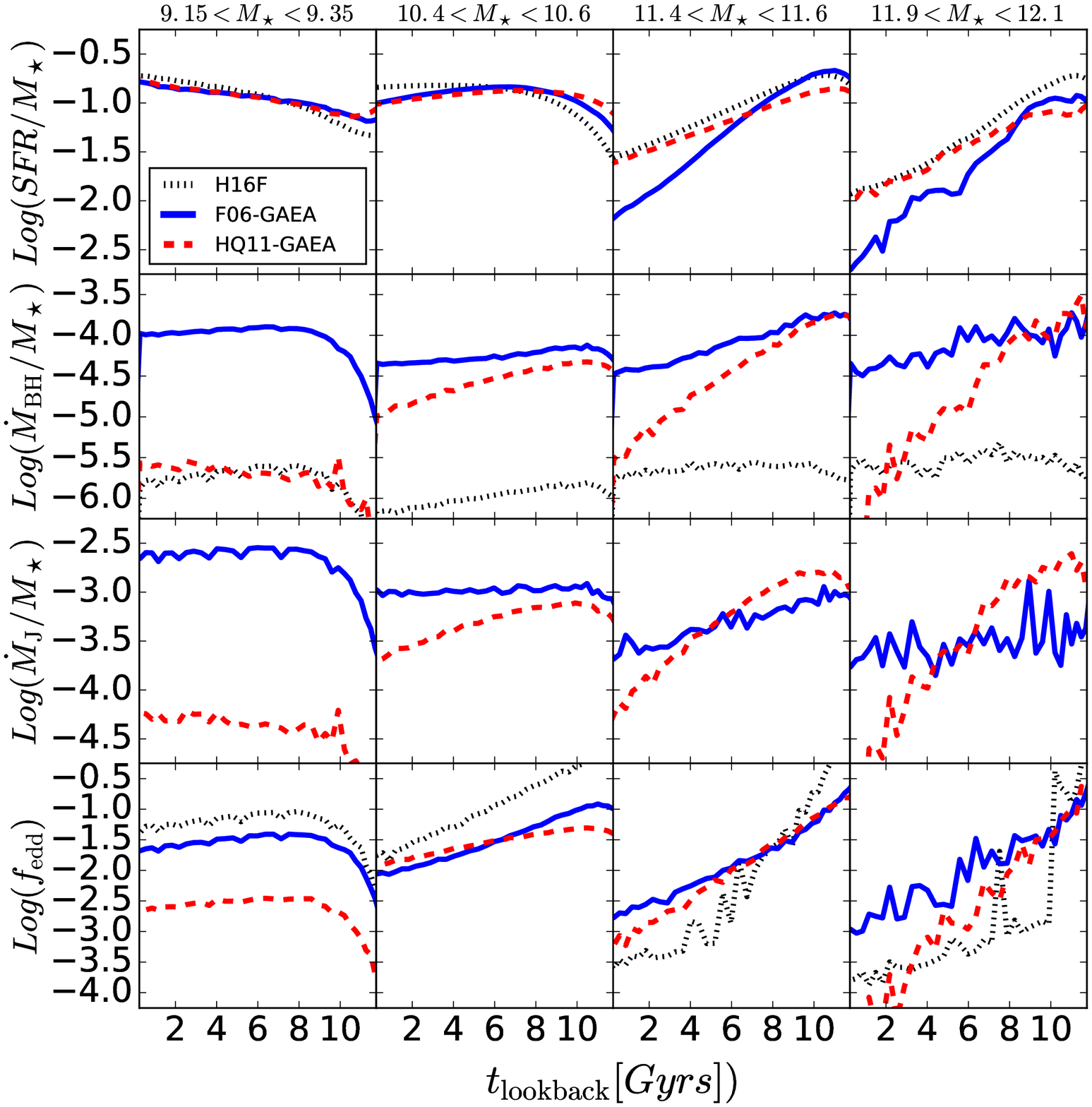} }
  \caption{Redshift evolution of the mean SFR normalized to
    $M_\star(z=0)$ (upper row), mean BH accretion rate normalized to
    $M_\star(z=0)$ (mid row) and mean $f_{\rm edd}$ for galaxies of
    different $z=0$ stellar mass (each column correspond to a
    different mass interval as indicated in the top label). Black,
    red and blue line refer to the different BH accretion
    models as in Fig.~\ref{fig:agnlf}.}\label{fig:histflu}
\end{figure*}
\begin{figure*}
  \centerline{ \includegraphics[width=18cm]{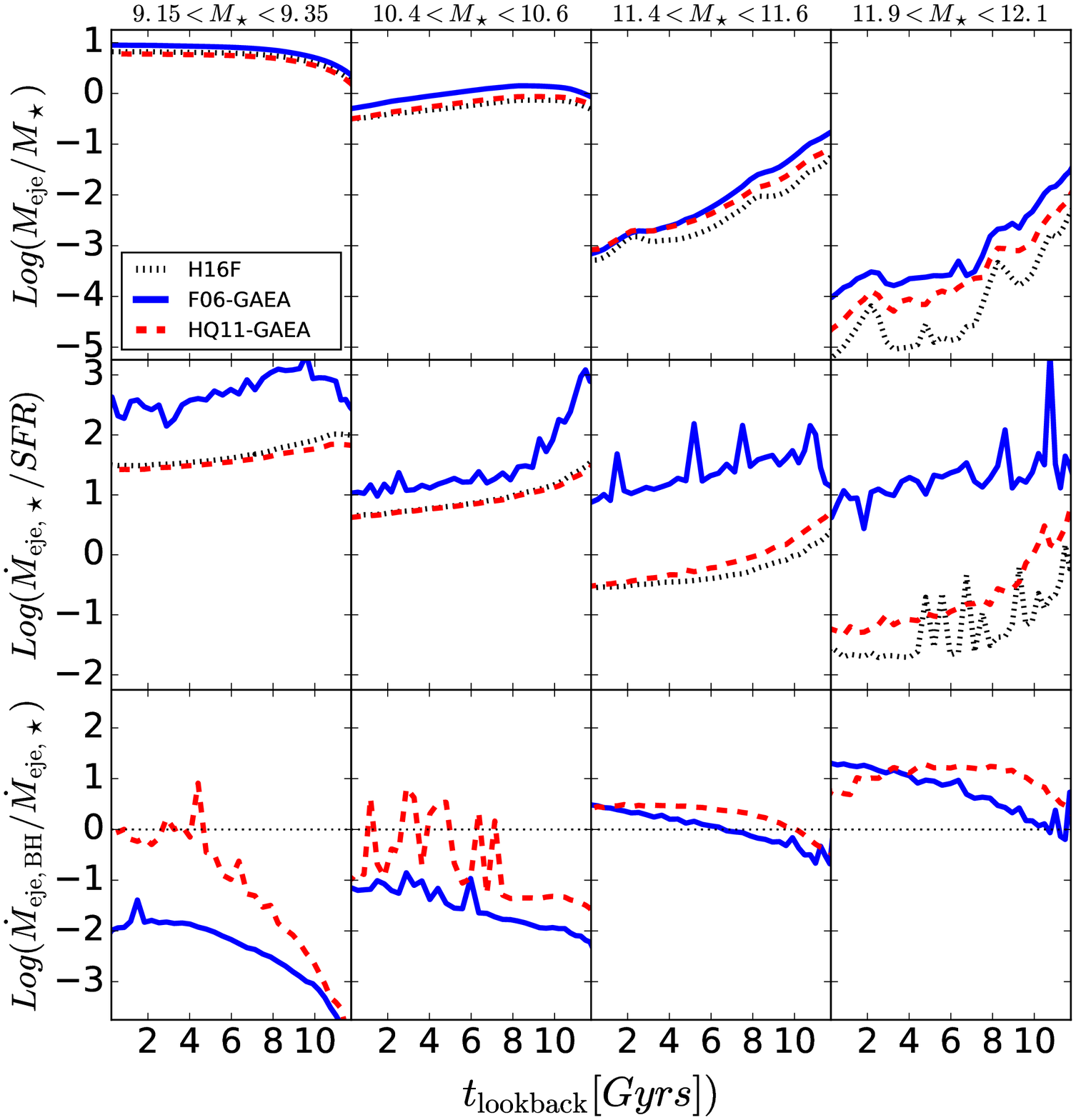} }
  \caption{Redshift evolution of the mass ejection rate in our
    realizations. The upper row shows the total ejected mass
    (normalized to $M_\star(z=0)$. The middle row shows the total mass
    loading factor (stellar plus AGN-driven winds), while the lower
    rows shows the ratio between the AGN-driven and stellar-driven
    ejection. Different columns refer to different present-day stellar
    mass as in Fig.~\ref{fig:histflu}. Black, red and blue lines
    refer to the different BH accretion models as in
    Fig.~\ref{fig:agnlf}.}\label{fig:histeje}
\end{figure*}
The two BH accretion models we implemented in {\gaea} provide
predictions for the basic properties of AGN that are in better
agreement with respect to the standard AGN model implemented in
H16F. This improvement is due to a number of reasons. First of all,
the new models include a delayed accretion onto SMBHs with respect to
previous model. This delay is due in {\ragaeaH} to the modelling of
the AGN light curve connected to each destabilising event, and in
{\ragaeaF} to the explicit modelling of the reservoir/torus around the
central SMBH (the viscous accretion timescale from Eq.~\ref{eq:bhaf06}
is equivalent to a light curve). The removal of the assumption of
instantaneous accretion in favour of a light curve approach has been
shown in previous work to improve the modelling of the faint-end of
the AGN-LF \citep{Marulli08, Lapi06}. Moreover, our new models provide
a much better description of the evolution of bright sources in
several respects. We implement an improved modelling for estimating
the amount of gas losing enough angular momentum to infall to the very
centre of the host galaxy. Another key aspect of the modelling lies in
the fact that we accumulate this low$-J$ material in a
reservoir/accretion disc, from where it can be accreted onto the
central SMBH on a non-instantaneous timescale. Finally, our approach
does not link BH accretion to merger events only, thus extending the
range of AGN triggering events.

There are some key differences between the two accretion schemes we
consider in this paper. Both approaches model the required loss of
angular momentum in the cold gas as a local process triggered by
``external'' events that happen on the scale of the host galaxy. In
both models, cold gas fuels the SMBH as an effect of gravitational
instabilities connected to large scale effects in the host disc
\citep[see e.g.][]{Romeo18}. In {\ragaeaH}, the amount of cold gas
available for accretion depends on the properties of the host galaxy,
while {\ragaeaF} assumes that the relevant mechanism is related to the
amount of SFR triggered in the central region of the galaxy. This
implies that in the {\ragaeaH} model, a disc instability event is
equivalent to a major merger, at fixed host galaxy properties, while
in {\ragaeaF} each triggering event is different from each other,
depending on the amount of SFR associated with it. Finally, it is
worth noticing that {\gaea} does not model galaxy interactions
(i.e. flybys, see e.g. \citealt{Menci08}), so that we neglect this
possible extra channel for AGN triggering, as well as direct smooth
gas accretion from the halo gas.

In order to better understand the effect of the considered BH
accretion schemes in {\gaea}, we show in Fig.~\ref{fig:histflu} and
~\ref{fig:histeje} the evolution of selected physical galaxy
properties. We consider galaxies in four representative mass bins
($Log(M_\star/\msun) \sim 9.25 - 10.5 - 11.5 - 12$) corresponding to
the four columns. In the upper panels of Fig.~\ref{fig:histflu} we
show how the BH accretion schemes modify the evolution of the sSFR. In
the largest mass bin, the {\ragaeaF} model is effective preferentially
at later times, while the {\ragaeaH} model impacts the sSFR already at
early cosmic epochs. In the second row, we consider the specific BH
accretion rate: this quantity is almost always larger in our improved
schemes with respect to KH00, with {\ragaeaF} always predicting larger
values than {\ragaeaH}. These trends are mainly due to the larger
amounts of cold gas typically infalling onto the reservoirs around the
SMBHs predicted by Eq.~\ref{eq:f06jloss} with respect to the
Eq.~\ref{eq:hq11jloss}, as shown in the third row. The increase of
cold gas available for accretion is particularly evident at low
redshift and for low-mass galaxies. Finally, in the lower panels we
show the evolution of the mean Eddington ratios. Although there is a
large population of SMBHs accreting at and above Eddington (see
Sec.~\ref{sec:eddlf}), the mean accretion takes place at sub-Eddington
rates.

Finally, we focus on the outflow rates associated with AGN-driven
winds, and we compare them to stellar-driven winds
(Fig.~\ref{fig:histeje}). The upper panels show the mean total
(i.e. AGN plus stellar driven winds) ejected mass, normalized to the
stellar mass in our reference mass bins. As expected, this quantity
increases in our new runs with respect to the standard {\gaea}
realization. This increase grows with the stellar mass of the galaxies
({\ragaeaF} systematically predicts larger ejected fractions than
{\ragaeaH}). A similar behaviour is also seen in the total mass
loading factor (middle panels): in this case the enhancement predicted
by the {\ragaeaF} run is particularly evident, while the {\ragaeaH}
run is closer to the standard run. The enhancement in {\ragaeaF} is
also due to the lower SFR predicted in this realization with respect
to H16F and {\ragaeaH}. Finally, in the lower panels, we directly
compare the ejection rates for the stellar-driven ($\dot{M}_{\rm eje,
  \star}$) and for the AGN-driven outflows ($\dot{M}_{\rm eje,
  BH}$). Our results show that the relative importance of these two
channels seems to correlate well with the final galaxy mass. This fact
highlights the importance of the different timescales associated with
the assembly of stellar mass and BH accretion in regulating their
relative evolution. Indeed, in more massive galaxies AGN-driven winds
systematically displace larger gas masses than stellar-driven winds at
all cosmic epochs. Their contribution is comparable at $M_\star \sim 3
\times 10^{11} \msun$. At lower stellar masses stellar-driven winds
still represent the key process to regulate the baryonic cycle, with
AGN-driven winds possibly relevant only at late times. This result is
consistent with our findings of a good agreement in the overall galaxy
properties between our new runs and the standard model: stellar-driven
winds are still the main regulating process for the evolution of
galaxies at the low-mass end of the GSMF.

The statistical properties of the AGN population and their evolution
have been analysed for different large-scale cosmological
hydro-dynamical simulations, including EAGLE \citep{RosasGuevara16},
IllustrisTNG \citep{Habouzit19a}, HorizonAGN \citep{Volonteri16},
Simba \citep{Dave19} and Magneticum \citep{Hirschmann14}. Due to the
limited resolution and to the lack of a complete theory describing the
physical processes under consideration, SMBH growth and the
corresponding feedback cannot be modelled from first
principles. Therefore sub-grid or sub-resolution models, including
free parameters, have to be implemented. In general, SMBHs of a given
mass are typically seeded in halos/galaxies above a certain mass
limit. These SMBH seeds are assumed to further grow via mergers with
other SMBHs and via gas accretion, mostly following the Bondi-Hoyle
approach. Feedback from SMBHs is modelled in a rather different way in
different simulations.

Despite the large variety of prescriptions adopted for BH growth and
feedback, and the uncertainties involved, simulations are to be fairly
successful, with minor short-comings, in reproducing basic,
statistical BH and AGN properties. The AGN feedback efficiency
parameter is typically calibrated to have the normalisation of the
$M_{\rm BH}$-$M_{\rm bul}$ relation consistent with observations. Most
simulations can also predict a fairly realistic evolution of AGN
luminosity functions. Some tensions remain, such as a slight
over-estimation of faint AGN at high redshifts in IllustrisTNG, or a
slight underestimation of bright AGN at $z\sim2$ in EAGLE. These
modern cosmological simulations have been tuned to match,
simultaneously, the evolution of the galaxy stellar mass
function. They are also able to roughly reproduce the quiescent
fractions of galaxies at a given stellar mass \citep[see][]{Xie20}.
The detailed colour distribution predicted for Illustris TNG has been
discussed in \citet{Nelson18}, while for the other simulations
discussed here, most comparison work has focused on the main sequence
of star-forming galaxies \citep{Sparre15, Katsianis17}. It is worth
stressing that hydro-simulations also report tensions in the balance
between the quenched fraction of massive and low-mass galaxies when
AGN feedback is taken into account, similar to our findings in
Fig.~\ref{fig:sfrhisto}. These results have motivated the
implementation of phenomenological prescriptions, in order to better
reproduce the observational constraints, such as mass-dependent
feedback schemes, either in the form of mass-dependent AGN feedback
schemes \citep{Dave19} or mass-dependent Eddington-limit for
quasar-mode, \citep{Weinberger18}.

A detailed comparison between results fromq these modern cosmological
simulations and predictions from {\gaea} is limited by the different
basic assumptions and modelling techniques. Models for AGN evolution
have been developed in the framework of the {\sc galform} SAM
\citep{Fanidakis12} and of the SantaCruz SAM
\citep{Hirschmann12}. Consistently with \citet{Hirschmann12}, we find
that disc instabilities are an important ingredient in the evolution
of the AGN population, but not the major contributors at all
luminosities, as found in \citet{Fanidakis12} model. In order to
reproduce the high-z AGN-LF and its downsizing trend,
\citet{Hirschmann12} use larger SMBH seeds than {\gaea}: this is
connected to the assumption of a strict Eddington limit for cold gas
accretion. As in \citet{Fanidakis12}, {\gaea} allows super-Eddington
accretion and thus requires a more moderate SMBH seeding to reproduce
the high-z evolution of the LF. However, while the downsizing trend in
the model by \citet{Fanidakis12} can be reproduced only assuming
relevant dust obscuration, this is not the case in {\gaea} and in
\citet{Hirschmann12}.

\section{Conclusions}\label{sec:final}
In this paper, we present an improved version of the {\gaea}
semi-analytic model featuring a state-of-the-art modelling of BH
accretion and AGN-driven winds. We model the AGN phenomenon
considering three different phases: (a) we first model the loss of
angular momentum required for the cold gas to fall towards the central
regions of the host galaxy and form a gas reservoir around its SMBH,
(b) we then estimate the accretion rate from the reservoir onto the
SMBH and (c) we study the effect of AGN-driven outflows on the
properties of the host galaxy. For each of the three phases we
consider two alternative prescriptions.

A first realization ({\ragaeaF}) is based on an analytic approach
similar to that adopted in the {\morgana} semi-analytic model
\citep{Fontanot06}, and inspired to the work of \citet{Umemura00}
and~\citet{Granato04}. This modelling assumes that the cold gas loses
angular momentum through a variety of processes that are linked to the
SFR of the host galaxies (i.e. turbulence and/or radiation drag). The
gas is accreted on the SMBH on a viscous timescale. The effect of the
AGN-driven winds is modelled using empirical prescriptions based on
observed AGN molecular and ionized outflows \citep{Fiore17}. As an
alternative model ({\ragaeaH}), we consider the $J$-loss rate
predicted by \citet{HopkinsQuataert11} and based on results from
high-resolution numerical simulations. These are aimed at studying the
back-reaction of gaseous star-forming discs whenever they are subject
to an instability (either due to the properties of the disc itself, or
induced by an interaction with a nearby object). We couple these
prescriptions with a light-curve model to estimate the accretion onto
the SMBH and with an analytic estimate of the ejection rate by AGN
proposed by \citet{Menci19}.

Among these two models, {\ragaeaF} shows better convergence properties
with respect to {\ragaeaH} against changes of the particle resolution
of the underlying simulation (see App.~\ref{app:resol}). Therefore, we
can use the {\ragaeaF} model to explore the properties of the AGN
population down to fainter magnitudes and smaller $M_{\rm BH}$. In
particular, the HQ11-GAEA model provides robust predictions at $z>1$
only for luminosities of the order of $L_\star$ or brighter
(i.e. around the knee of the AGN-LF), while {\ragaeaF} extends to
$\sim 0.01 L_\star$ up to $z \sim 4$. This behaviour is connected to
the modelling of $J-$loss rates following
\citet{HopkinsQuataert11}. This description relies on the structural
properties of the inner regions of simulated galaxies, which become
increasingly difficult to recover at increasing redshift. In order to
explore the AGN properties for sources below the knee of the LF at
$z\sim2$, a different simulation is needed with a better resolution
than MS and comparable cosmological volumes.

We stress that the inclusion of an AGN phase has only a marginal
effect on the overall assembly of the galaxy population, as seen from
the evolution of the GSMF. However, the effect of AGN-driven winds on
the SFR of massive galaxies can be relevant, helping in displacing
larger amounts of cold gas with respect to stellar feedback driven
winds. Our results clearly support a scenario where the {\it combined}
effect of AGN and stellar feedback is fundamental in order to
reproduce the observed AGN and host galaxy properties {\it at the same
  time}. In particular, AGN-driven winds help in reproducing the (low)
levels of SFR in massive galaxies, removing some of the cold gas still
in place in these galaxies since $z\sim2$. However, the removal of gas
is strong in low-mass galaxies as well, worsening the agreement with
observational constraints found in H16F for the passive fraction at
$M_\star<10^{10} \msun$. Overall these results suggest that AGN-driven
winds alone cannot be the solution for all problems highlighted in
H16F. A deeper revision of the stellar feedback modelling is required
as well, possibly taking into account the coupling between stellar and
AGN feedback \citep{MonacoFontanot05}, or assuming a mass-dependent
feedback \citep[see e.g.][]{Dave19}.

Another interesting aspect lies in the predicted Eddington rates. Our
results show clearly that the more massive SMBHs (i.e. $M_{\rm BH}
\gtrsim 10^8 \msun$) are already in a self-regulated regime, able to
reproduce the observed distribution of Eddington ratios at various
redshifts (that implies a decreasing number of sources at increasing
Eddington ratio). Smaller SMBHs have not reached this stage yet, and
show a rather different distribution of Eddington ratios, flattening
at high $f_{\rm edd}$ and implying that all possible accretion rates
are plausible. This is due to the availability of large gas reservoirs
in the host galaxy, that smaller central objects are not able to
displace efficiently as more massive systems. Consistently, our models
predict a large contribution of AGN powered by small SMBHs
(i.e. $M_{\rm BH}<10^6 \msun$) to the space density of AGN around the
knee of the LF. Testing this prediction is beyond the capabilities of
current instrumentation, but future facilities, like Athena, hold the
promise to provide the required insight.

Finally, we show that the assumption that mechanisms other than galaxy
mergers can trigger an AGN event as well as the inclusion of a delayed
accretion model for the cold gas (either in the form of a gas
reservoir or as a light curve) are important to reproduce the overall
shape of the AGN LF. This implies that a detailed treatment of disc
instabilities is critical for reproducing the AGN
population. \citet{DeLucia11} showed that disc instability are a
fundamental process for bulge growth at intermediate host galaxy
masses (i.e. $10^{10} < M_\star/\msun < 10^{11} $). The most relevant
difference between our realizations lies in the treatment of disc
instabilities: their impact is larger in the {\ragaeaH} realization
because this model assumes that the amount of $J$-loss in the cold gas
depends on the local properties of the host galaxy, i.e. on the
stellar and cold gas mass distributions, independently on magnitude of
the mass transfer involved. In {\ragaeaF}, instead, the impact of disc
instabilities is proportional to the amount of stellar mass moved from
the disc to the bulge (to get the disc back to stability). The actual
implementation of disc instabilities in {\gaea}, i.e. transferring
from the disc to the bulge just the amount of stars required to
restore disc stability, is rather conservative and possibly too
simplistic, as it predicts very frequent small mass transfers. These
correspond to small low-J gas flows in {\ragaeaF}, but too many disc
instability episodes in {\ragaeaH}.  The relative importance of disc
instabilities on the galaxy evolution is still a long standing issue
in many theoretical models \citep[see e.g.][]{DeLucia11}: alternative
approaches have been used in different galaxy evolution models but it
is currently unclear which is the most realistic way to model the fate
of an unstable disc (i.e. the corresponding mass and energy
transfers).

BH accretion and AGN-driven outflows represent a key ingredient in
modern models of galaxy formation and evolution. These models provide
relevant insight for the physical interpretation of the observed
frequency and properties of gaseous outflows, in and around AGN host
galaxies, that have become available thanks to the advent of
instruments, like the Atacama Large Millimeter Array (ALMA) and the
Multi Unit Spectroscopic Explorer (MUSE) on the Very Large Telescope
(VLT). These, together with next generation instruments like the
Enhanced Resolution Imager and Spectrograph (ERIS) on VLT, allow
detailed spectro-imaging of multiple gas components, down to pc scale
in nearby galaxies, and {\rm kpc} scale at redshift of $\sim 2$, and
revealed the ubiquitous nature of the outflows in AGN host galaxies
\citep[see e.g.][]{Shimizu19, ForsterSchreiber19, Feruglio20}.

\section*{Acknowledgements}
FF thanks Angela Bongiorno, Anna Gallazzi, Emanuele Giallongo,
Elisabeta Lusso, Andrea Merloni, Alex Saro and Stefano Zibetti for
useful discussions and their help in a better understanding of the
different facets of data analysis. FF, NM, FF and CF acknowledge
support from PRIN MIUR project ``Black Hole winds and the Baryon Life
Cycle of Galaxies: the stone-guest at the galaxy evolution supper'',
contract 2017-PH3WAT. MH acknowlegdes financial support from the
Carlsberg Foundation via a ``Semper Ardens'' grant (CF15-0384).FS
acknowledges partial support from a Leverhulme Trust Research
Fellowship. We also acknowledge the computing centre of INAF-OATs,
under the coordination of the CHIPP project \citep{Taffoni20}, for the
availability of computing resources and support.

\section*{Data Availability}
An introduction to {\gaea}, a list of our recent work, as well as
datafile containing published model predictions, can be found at
\url{http://adlibitum.oats.inaf.it/delucia/GAEA/}

\bibliographystyle{mnras}
\bibliography{fontanot}

\newpage
\appendix

\section{Effect on global properties of galaxy populations}\label{app:b}
\begin{figure}
  \centerline{ \includegraphics[width=9cm]{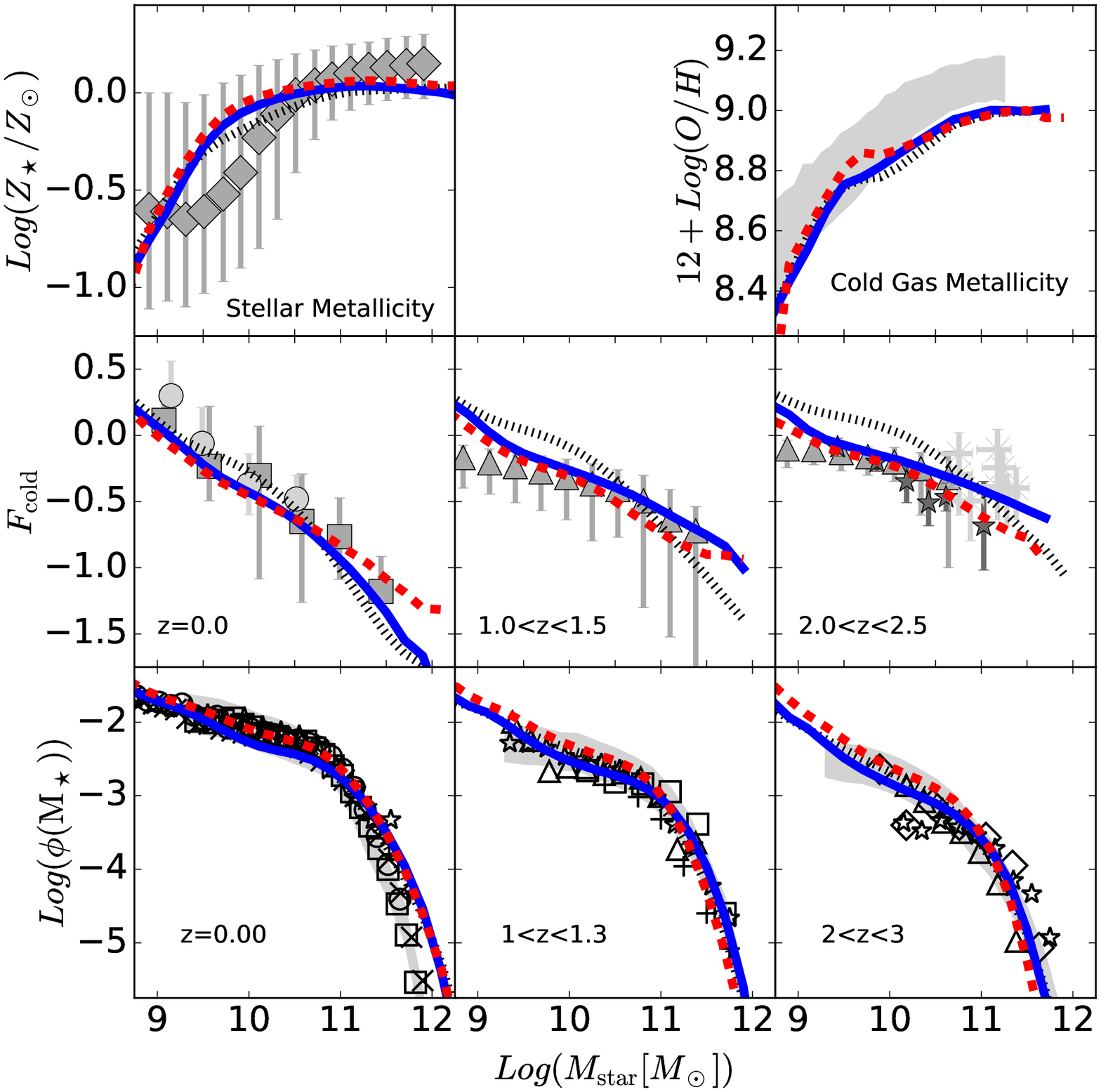} }
  \caption{Physical properties of galaxy population: comparison with
    previous results. {\it Upper panels:} Stellar and cold gas
    mass-Metallicity relations at $z\sim0$; {\it middle panels:} redshift
    evolution cold gas fractions in star forming galaxies; {\it lower
      panels:} redshift evolution of the galaxy stellar mass
    function. Data as in \citet{Hirschmann16}. In all panels, black,
    red and blue lines refer to the different BH accretion models as
    in Fig.~\ref{fig:agnlf}.}\label{fig:appevo}
\end{figure}
\begin{figure}
  \centerline{ \includegraphics[width=9cm]{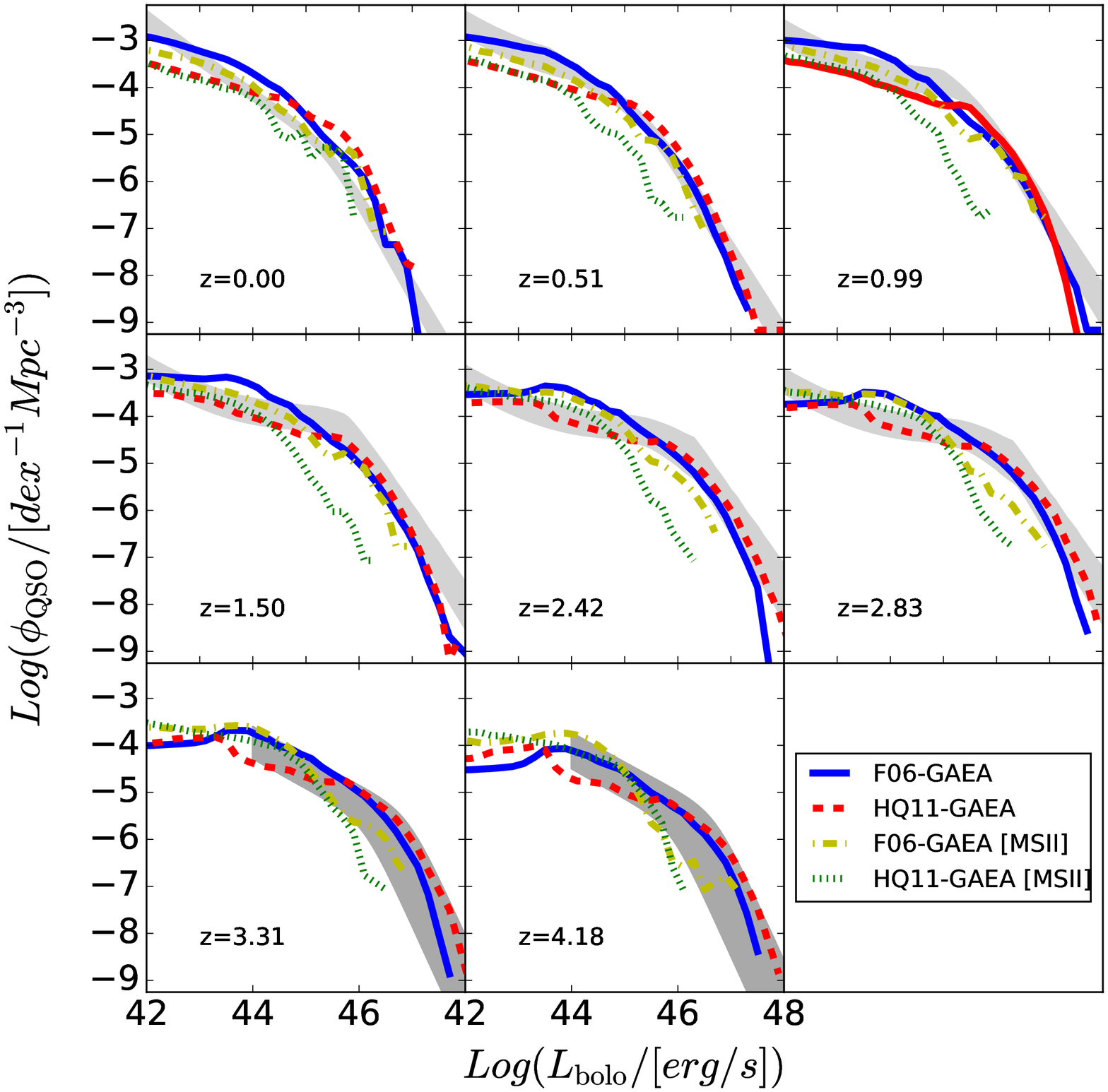}}
  \caption{Redshift evolution of the galaxy stellar mass function:
    comparison between MS reference runs (red and blue lines for
    {\ragaeaF} and {\ragaeaH} respectively) and realizations using
    MSII (green and yellow lines for {\ragaeaF} and {\ragaeaH}
    respectively). Data as in Fig.~\ref{fig:agnlf}.}\label{fig:appresol}
\end{figure}
In this appendix we briefly discuss the effect of the different BH
accretion schemes on the overall galaxy properties with respect to the
standard H16 run. Fig~\ref{fig:appevo} shows that the main predictions of
the H16 model are robust against the inclusion of improved BH
accretion prescription and the QSO mode feedback. The main difference
we find with respect to H16 predictions lies in the evolution of the
amount of cold gas available and sSFR distributions as AGN-driven
winds are able to displace a larger amount of material with respect to
stellar-driven winds alone.

\section{Resolution Effects}\label{app:resol}
In Fig.~\ref{fig:appresol} we show predictions from our model
realizations run on the MSII with the same parameters calibrated on
the Millennium Simulation. The MSII represents a numerical experiment
with the same cosmological parameters as the MS, but its smaller
volume ($100^3 {\rm Mpc}^3$ instead of $500^3 {\rm Mpc}^3$) allows to
resolve smaller structures on the same numerical grid (the MSII has a
resolution 125 times better than the MS). Fig.~\ref{fig:appresol}
shows that neither of the models presented in this paper achieves a
good level of convergence at the faint-end of the AGN-LF. We use these
results to estimate for each model a redshift-dependent luminosity
limit above which we consider model predictions robust. This limit
corresponds to the transition from solid to dashed line in
Fig.~\ref{fig:agnlf}.

\section{Alternative combinations of prescriptions}\label{app:multimod}
\begin{figure*}
  \centerline{ \includegraphics[width=9cm]{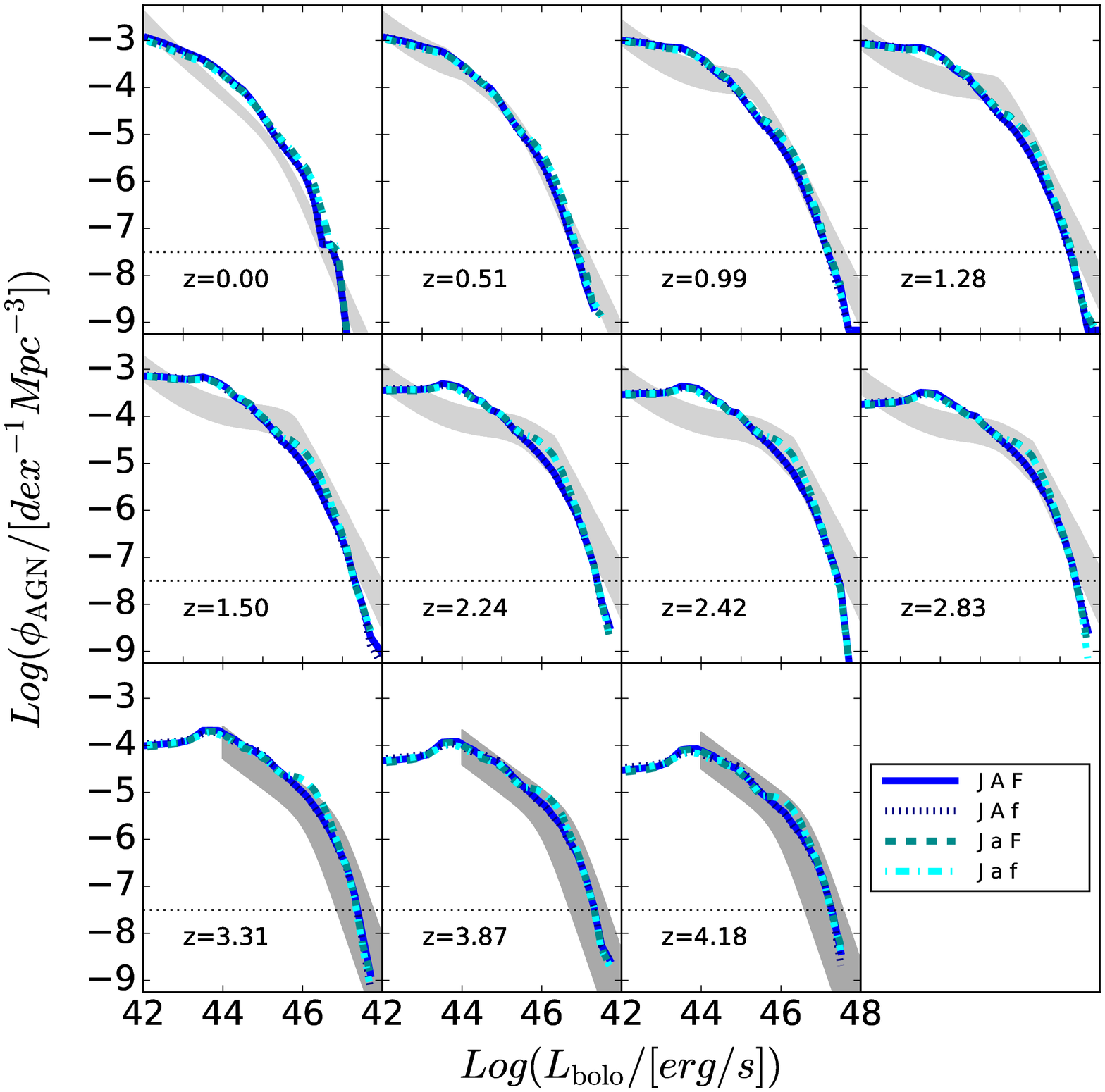}
    \includegraphics[width=9cm]{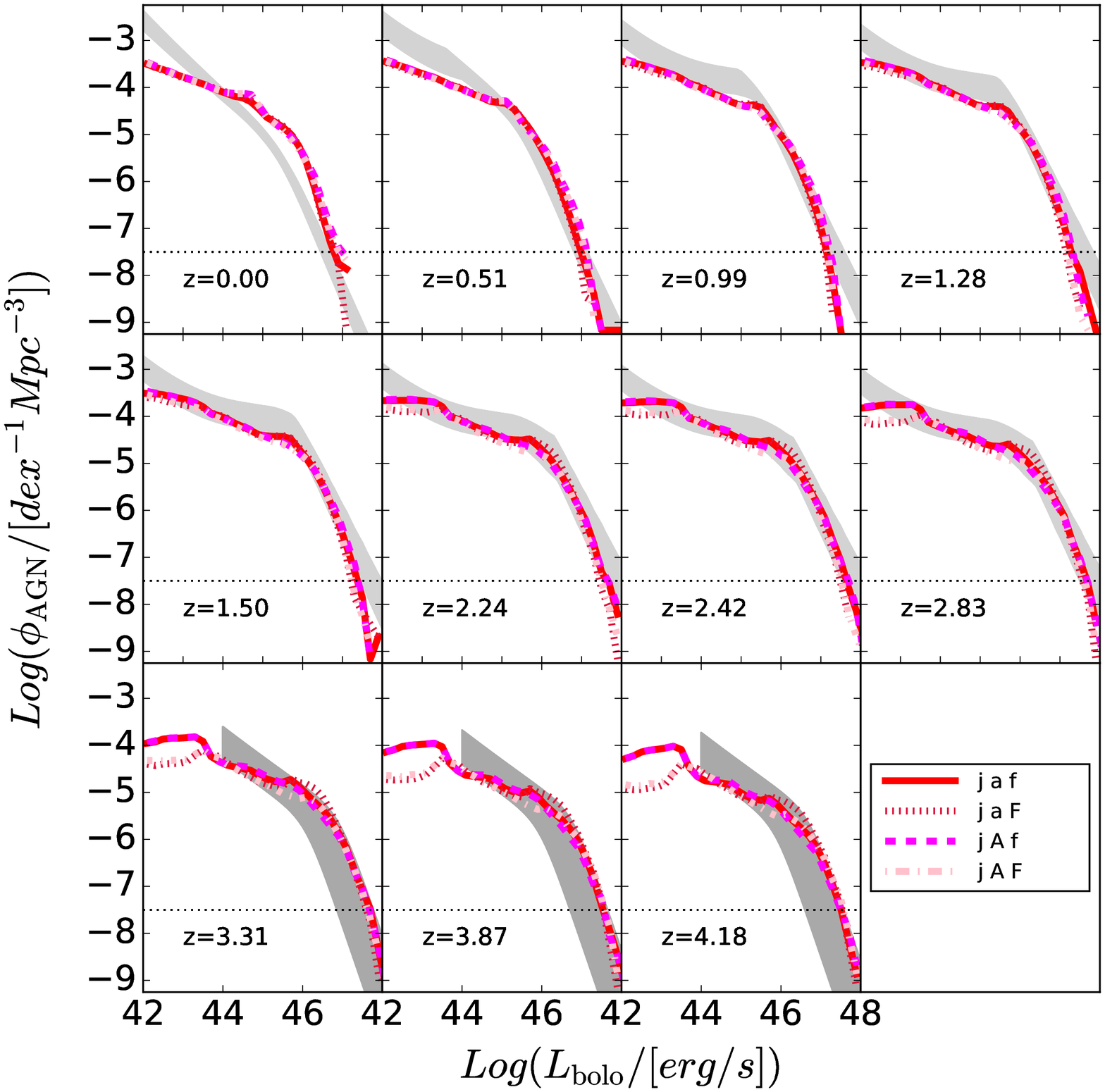} }
  \caption{Redshift evolution of the of the AGN LF for alternative
    prescription combinations. Linestyle and colors refer to different
    runs as labelled. For each prescription, capital and lower case
    letters refer to the first and second choice in the present
    manuscript. Data as in Fig.~\ref{fig:agnlf}.}\label{fig:multimod}
\end{figure*}

The different prescriptions we define in Sec.~\ref{sec:models} can be
combined in 8 different models. In the main paper we focus on just two
of these prescriptions, and in this appendix we show that these are
representative for all different choices. In the two panels of
Fig.~\ref{fig:multimod} we show all 8 models. We label them using
capital and lower case letters, that refer to the first and second
choice in the present manuscript, respectively. ``J'' or ``j'' refer
to the $J$-loss prescriptions (sec.~\ref{sec:inflowA}
and~\ref{sec:inflowB}); ``A'' and ``a'' to the accretion prescriptions
(sec.~\ref{sec:accrA} and~\ref{sec:accrB}); ``F'' and ``f'' to the
outflow prescription (sec.~\ref{sec:outflowA}
and~\ref{sec:outflowB}). Within this convention {\ragaeaF} and
{\ragaeaH} correspond to the ``J A F'' and ``j a f'' combinations,
respectively. Moreover, for all model variants we consider the same
reference values of the parameters as in table~\ref{tab:parameters},
to highlight the effect of the different combinations. The two panels
in Fig.~\ref{fig:multimod} contain 4 models each, keeping constant the
$J$-loss prescription: it's quite evident that this is the
prescription that provides the largest impact on the model predictions
for the AGN population.

\end{document}

%% file: journal.tex
\def\aj{AJ}%
\def\araa{ARA\&A}%
\def\apj{ApJ}%
\def\apjl{ApJ}%
\def\apjs{ApJS}%
\def\aap{A\&A}%
\def\mnras{MNRAS}%
\def\nat{Nature}%
